\documentclass[12pt]{article}
\usepackage{graphicx}

\begin{document}

\title{Decoherence of the quantum logic gate implemented with the Jaynes-Cummings model: A semiclassical approach}

\author{Hiroo Azuma\thanks{Email: hiroo.azuma@m3.dion.ne.jp}
\\
{\small Advanced Algorithm \& Systems Co., Ltd.,}\\
{\small 7F Ebisu-IS Building, 1-13-6 Ebisu, Shibuya-ku, Tokyo 150-0013, Japan}
}

\date{\today}

\maketitle

\begin{abstract}
In this paper,
we investigate decoherence of Knill, Laflamme, and Milburn's nonlinear sign-shift gate
that is implemented with the Jaynes-Cummings model.
Introducing a stochastic variable as an external electric field,
we let it couple with a dipole moment of a two-level atom.
We examine this model using a semiclassical theory.
Results of the Monte Carlo simulations under the semiclassical approximation correspond well
with those obtained with the quantum mechanical perturbation theory
for the stochastic process.
In the results of the simulations,
we observe both the $\mbox{T}_{1}$ and $\mbox{T}_{2}$ decays.
This paper is a sequel to the reference
[H.~Azuma, Prog. Theor. Phys. {\bf 126}, 369 (2011)].
\end{abstract}

\section{\label{section-introduction}Introduction}
Since Shor's and Grover's quantum algorithms were discovered,
about twenty years have passed \cite{Shor1997,Grover1997}.
However, we have not built a stable and scalable quantum computer yet.
Thus,
realization of the quantum computer has become one of the most exciting research topics
for both theoretical and experimental physicists.

Shor's algorithm factorizes large integers more efficiently than classical algorithms.
We can consider it to be a serious threat to public-key cryptosystems.
Grover's algorithm can be viewed as an efficient amplitude-amplification process for quantum states.
Applying the same unitary transformation to a uniform superposition of basis vectors in successive iteration,
it amplifies an amplitude of a certain basis vector that an oracle indicates.

To construct a quantum computer,
we need to prepare quantum bits (qubits),
which are two-state quantum systems,
and quantum logic gates,
which apply unitary transformations to qubits.
In this paper,
as mentioned later,
we construct the qubit with a photon running on a pair of optical paths,
and we can apply an arbitrary $U(2)$ transformation to the single qubit
using beamsplitters and waveplates (phase shifters) with ease.
Contrastingly, many researchers consider implementation of a two-qubit gate to be very difficult
because it has to generate nonlocal quantum correlation,
which is called entanglement, between two qubits.
Moreover, it is shown that we can construct any unitary transformation applied to an arbitrary number of qubits out of one-qubit transformations
and certain two-qubit gates,
such as the controlled-NOT gates and the conditional sign-flip gates \cite{Barenco1995}.
Hence, many researchers concentrate on building two-qubit gates that generate entanglement.

So far, a lot of methods for implementing two-qubit quantum logic gates have been proposed.
In Refs.~\cite{Cirac1995,Monroe1995},
cold-trapped ion quantum computation is examined and demonstrated in the laboratory.
In Ref.~\cite{Turchette1995},
photons in the cavity quantum electrodynamics system are utilized as qubits.
The nuclear magnetic resonance quantum computer is proposed and demonstrated
in Refs.~\cite{Gershenfeld1997,Jones1998}.
In Ref.~\cite{Kane1998},
implementation of qubits with an array of the nuclear spins of phosphorus donor atoms fixed into a doped silicon lattice is proposed.
In Refs.~\cite{Briegel2001,Raussendorf2001,Walther2005},
the one-way quantum computer is proposed and experimentally realized.
In Ref.~\cite{Trauzettel2007},
spin qubits built with graphene quantum dots are discussed.

Knill, Laflamme, and Milburn (KLM) show an ingenious method for applying the conditional sign-flip gate to dual-rail qubits
using beamsplitters and the nonlinear sign-shift (NS) gates \cite{Knill2001,Ralph2001,Kok2007},
which cause the following transformation to the number states of photons:
\begin{equation}
\alpha|0\rangle_{\mbox{\scriptsize P}}
+
\beta|1\rangle_{\mbox{\scriptsize P}}
+
\gamma|2\rangle_{\mbox{\scriptsize P}}
\longrightarrow
\alpha|0\rangle_{\mbox{\scriptsize P}}
+
\beta|1\rangle_{\mbox{\scriptsize P}}
-
\gamma|2\rangle_{\mbox{\scriptsize P}}.
\label{definition-NS-gate-0}
\end{equation}
The index P stands for the photons.
In KLM's method,
we regard the pair of optical paths where the single photon is running as the qubit.
This construction of the qubit is called the dual-rail qubit representation \cite{Chuang1995}.

In Ref.~\cite{Azuma2011}, the author of the current paper shows a method for implementing the NS gate
with the Jaynes-Cummings model (JCM).
The JCM is a quantum mechanical model describing the interaction between a single two-level atom and a single electromagnetic field mode in a cavity.
It was originally designed for generating a spontaneous emission of the atom in 1963 \cite{Jaynes1963}.
As a typical soluble model for the cavity quantum electrodynamics,
comprehensive study of the JCM is made theoretically \cite{Shore1993,Louisell1973,Walls1994,Schleich2001}.
An experimental demonstration was performed in 1987 \cite{Rempe1987}.
Because of these achievements,
the JCM is very well-studied and familiar to the researchers in the field of quantum optics.
Thus, we can expect that this proposal is more practical and feasible than other proposals mentioned above.

In KLM's proposal, the NS gate is constructed only with passive linear optics,
and it works as a nondeterministic gate conditioned on the detection of an auxiliary photon.
It works with probability $1/4$.
In Ref.~\cite{Azuma2011},
we introduce a nonlinear device into KLM's scheme against KLM's original idea that the two-qubit gate can be constructed with only passive linear optics.
However, in our method,
the NS gate works with small error probability
and the author of the current paper thinks that our method is a practical alternative for the simplification of the whole system
of the quantum computer.

In Ref.~\cite{Gilchrist2003},
Gilchrist {\it et al}. try to build the NS gate by trapped atoms in an optical cavity.
In Ref.~\cite{Azuma2008},
the author of the current paper proposes a method of constructing the NS gate with a one-dimensional Kerr-nonlinear photonic crystal.
In Ref.~\cite{Lemr2010},
the NS gate is realized experimentally with linear optical components according to the original scheme of KLM's.

In this paper,
we investigate decoherence of KLM's NS gate that is implemented with the JCM \cite{Azuma2011}.
In general, decoherence is gradual loss of coherence of a quantum system
and it is given rise to by unexpected interaction between the quantum system and its external environment.
If we want to demonstrate our implementation in the laboratory,
we have to analyze its decoherence precisely.
This is because actual experiments of the JCM
are always disturbed by thermal effects and noisy external fields.
From practical viewpoints,
we cannot neglect these disturbances for a real experimental setup.
Because to examine decoherence in quantum logic gate is important,
there are a lot of prior works on this topic \cite{Thorwart2001,Barreiro2011,vanderSar2012}.

In this paper,
we analyze the decoherence by the following method.
Introducing a stochastic variable as an external electric field,
we let it couple with a dipole moment of the two-level atom.
We examine this model with a semiclassical theory.
Physical quantities are evaluated with the Monte Carlo simulations.
They correspond well with the results obtained with the quantum mechanical perturbation theory
for the stochastic process.
This semiclassical treatment is inspired by Ref.~\cite{Palma1996}.
(In Ref.~\cite{Palma1996},
the decoherence of quantum registers is investigated comprehensively.)
In results of the simulations,
we observe both the $\mbox{T}_{1}$ and $\mbox{T}_{2}$ decays.
This paper is a sequel to Ref.~\cite{Azuma2011}.

This paper is organized as follows.
In Sect.~\ref{section-construction-of-NS-gate-with-JCM},
we explain how to implement the NS gate with the JCM.
In Sect.~\ref{section-semiclassical-decoherence-model},
we introduce a semiclassical model that describes the decoherence of the NS gate.
In Sect.~\ref{section-numerical-simulation-semiclassical-model},
we show results of numerical simulations for the semiclassical model.
In Sect.~\ref{section-perturbation-theory-fidelity},
we examine time variation of a fidelity of the NS gate for the semiclassical model
using the quantum mechanical perturbation theory for the stochastic process.
In Sect.~\ref{section-discussion},
we give brief discussion.
In Appendix~\ref{appendix-electric-field-dipole-interaction},
we explain the electric field-dipole interaction.
In Appendix~\ref{appendix-stochastic-variable},
we
calculate variance and a distribution of the stochastic variable that is used in the semiclassical model
as the external field.

\section{\label{section-construction-of-NS-gate-with-JCM}Implementation of the NS gate with the JCM}
In this section,
we explain how to implement the NS gate with the JCM.
This section is a brief review of Ref.~\cite{Azuma2011}.
First of all, we assume that the field is resonant with the atom,
and the photons' frequency times Planck's constant is equal to the energy gap of the two-level atom.
Then, we write the JCM's Hamiltonian as
\begin{eqnarray}
H&=&H_{0}+H_{\mbox{\scriptsize I}}, \nonumber \\
H_{0}&=&\hbar\omega[(1/2)\sigma_{z}+a^{\dagger}a], \nonumber \\
H_{\mbox{\scriptsize I}}&=&\hbar g(\sigma_{+}a+\sigma_{-}a^{\dagger}),
\label{Hamiltonian-definition-0}
\end{eqnarray}
where
$\sigma_{\pm}=(1/2)(\sigma_{x}\pm i\sigma_{y})$
and
$[a,a^{\dagger}]=1$.
The Pauli matrices
$\{\sigma_{i}:i=x,y,z\}$
are operators acting on the atom,
and
$a$
and
$a^{\dagger}$
are the annihilation and creation operators acting on the electromagnetic field, respectively.
Here, we assume that $g$ is a real constant.

Because
$[H_{0},H_{\mbox{\scriptsize I}}]=0$
and
we can diagonalize
$H_{0}$
with ease,
we take the following interaction picture.
We describe a state vector of the whole system in the Schr\"{o}dinger picture as
$|\psi_{\mbox{\scriptsize S}}(t)\rangle$.
We define a state vector in the interaction picture as
$|\psi_{\mbox{\scriptsize I}}(t)\rangle=\exp(iH_{0}t/\hbar)|\psi_{\mbox{\scriptsize S}}(t)\rangle$
with assuming
$|\psi_{\mbox{\scriptsize I}}(0)\rangle=|\psi_{\mbox{\scriptsize S}}(0)\rangle$.
The time evolution of
$|\psi_{\mbox{\scriptsize I}}(t)\rangle$
is given by
$|\psi_{\mbox{\scriptsize I}}(t)\rangle=U_{\mbox{\scriptsize I}}(t)|\psi_{\mbox{\scriptsize I}}(0)\rangle$,
where
$U_{\mbox{\scriptsize I}}(t)=\exp(-iH_{\mbox{\scriptsize I}}t/\hbar)$.

We define the basis vectors for the state of the atom and the photons as follows.
The ground and excited states of the atom are given by two-component vectors,
\begin{equation}
|g\rangle_{\mbox{\scriptsize A}}
=
\left(
\begin{array}{c}
0 \\
1 \\
\end{array}
\right),
\quad\quad
|e\rangle_{\mbox{\scriptsize A}}
=
\left(
\begin{array}{c}
1 \\
0 \\
\end{array}
\right),
\end{equation}
respectively.
The index A stands for the atom.
The number states of the photons are given by
$|n\rangle_{\mbox{\scriptsize P}}$,
where
$n=0, 1, 2, ...$.
Describing the atom's Pauli operators with
$2\times 2$
matrices,
we can write down
$U_{\mbox{\scriptsize I}}(t)$
as follows:
\begin{equation}
U_{\mbox{\scriptsize I}}(t)
=
\exp[-it
\left(
\begin{array}{cc}
0 & ga \\
ga^{\dagger} & 0 \\
\end{array}
\right)
]
=
\left(
\begin{array}{cc}
u_{00} & u_{01} \\
u_{10} & u_{11} \\
\end{array}
\right),
\end{equation}
where
\begin{eqnarray}
u_{00}
&=&
\cos(|g|\sqrt{a^{\dagger}a+1}t), \nonumber \\
u_{01}
&=&
-
iga\frac{\sin(|g|\sqrt{a^{\dagger}a}t)}{|g|\sqrt{a^{\dagger}a}}, \nonumber \\
u_{10}
&=&
-
iga^{\dagger}\frac{\sin(|g|\sqrt{a^{\dagger}a+1}t)}{|g|\sqrt{a^{\dagger}a+1}}, \nonumber \\
u_{11}
&=&
\cos(|g|\sqrt{a^{\dagger}a}t).
\end{eqnarray}

In this paper,
we use two orthonormal bases.
The first one diagonalizes $H_{0}$
and it is given by
\begin{equation}
\left\{
\begin{array}{ll}
|g,n\rangle=|g\rangle_{\mbox{\scriptsize A}}|n\rangle_{\mbox{\scriptsize P}} & \mbox{for $n=0, 1, 2, ...$,} \\
|e,n\rangle=|e\rangle_{\mbox{\scriptsize A}}|n\rangle_{\mbox{\scriptsize P}} & \mbox{for $n=0, 1, 2, ...$.} \\
\end{array}
\right.
\label{eigen-system-1}
\end{equation}
The second one diagonalizes $H_{\mbox{\scriptsize I}}$
and it is given by
\begin{equation}
\{
|g,0\rangle,
|n_{\pm}\rangle=(1/\sqrt{2})(|e,n\rangle\pm|g,n+1\rangle)
\quad
\mbox{for $n=0, 1, 2, ...$}
\}.
\label{eigen-system-2-0}
\end{equation}
Eigenvalues of $H_{\mbox{\scriptsize I}}$ for $\{|g,0\rangle,|n_{\pm}\rangle : n=0,1,2,...\}$ are given as follows:
\begin{eqnarray}
H_{\mbox{\scriptsize I}}|g,0\rangle
&=&
0, \nonumber \\
H_{\mbox{\scriptsize I}}|n_{\pm}\rangle
&=&
E_{n_{\pm}}|n_{\pm}\rangle, \nonumber \\
E_{n_{\pm}}
&=&
\pm\hbar g\sqrt{n+1}
\quad
\mbox{for $n=0, 1, 2, ...$.}
\label{eigen-system-2-1}
\end{eqnarray}
The second orthonormal basis $\{|g,0\rangle,|n_{\pm}\rangle\}$ plays an important role
in Sects.~\ref{section-numerical-simulation-semiclassical-model}
and \ref{section-perturbation-theory-fidelity}.

We can write down the time evolution of the three initial states,
$|\psi_{\mbox{\scriptsize I}}(0)\rangle=|g,0\rangle$,
$|g,1\rangle$,
and
$|g,2\rangle$
as
\begin{eqnarray}
U_{\mbox{\scriptsize I}}(t)|g,0\rangle
&=&
U_{\mbox{\scriptsize I}}(t)
\left(
\begin{array}{c}
0 \\
|0\rangle_{\mbox{\scriptsize P}} \\
\end{array}
\right)
=
\left(
\begin{array}{c}
0 \\
|0\rangle_{\mbox{\scriptsize P}} \\
\end{array}
\right), \nonumber \\
U_{\mbox{\scriptsize I}}(t)|g,1\rangle
&=&
U_{\mbox{\scriptsize I}}(t)
\left(
\begin{array}{c}
0 \\
|1\rangle_{\mbox{\scriptsize P}} \\
\end{array}
\right)
=
\left(
\begin{array}{c}
-i(g/|g|)\sin(|g|t)|0\rangle_{\mbox{\scriptsize P}} \\
\cos(|g|t)|1\rangle_{\mbox{\scriptsize P}} \\
\end{array}
\right), \nonumber \\
U_{\mbox{\scriptsize I}}(t)|g,2\rangle
&=&
U_{\mbox{\scriptsize I}}(t)
\left(
\begin{array}{c}
0 \\
|2\rangle_{\mbox{\scriptsize P}} \\
\end{array}
\right)
=
\left(
\begin{array}{c}
-i(g/|g|)\sin(\sqrt{2}|g|t)|1\rangle_{\mbox{\scriptsize P}} \\
\cos(\sqrt{2}|g|t)|2\rangle_{\mbox{\scriptsize P}} \\
\end{array}
\right).
\end{eqnarray}
To obtain the NS gate,
we have to flip only the sign of the coefficient of
$|2\rangle_{\mbox{\scriptsize P}}$.
Thus, we let
$t=(2m+1)\pi/(\sqrt{2}|g|)$
for
$m=0, 1, 2, ...$,
and obtain the following time evolution:
$|g,0\rangle
\to
|g,0\rangle$,
$|g,1\rangle
\to
c(m)|e,0\rangle
+
d(m)|g,1\rangle$,
$|g,2\rangle
\to
-|g,2\rangle$,
where
$c(m)=-i(g/|g|)\sin[(2m+1)\pi/\sqrt{2}]$
and
$d(m)=\cos[(2m+1)\pi/\sqrt{2}]$.

\begin{table}
\caption{Variations of $|c(m)|^{2}$,
the error probability,
and $d(m)$,
the coefficient of $|g,1\rangle$ in the evolved state,
for $m=0, 1, 2, 3, 4$.}
\label{Table01}
\begin{center}
\begin{tabular}{|c|c|c|} \hline
$m$ & $|c(m)|^{2}$ & $d(m)$ \\
\hline
$0$ & $0.633$ & $-0.606$ \\
$1$ & $0.138$ & $0.928$  \\
$2$ & $0.988$ & $0.111$  \\
$3$ & $0.0247$ & $-0.988$ \\
$4$ & $0.828$ & $0.415$  \\
\hline
\end{tabular}
\end{center}
\end{table}

In Table~\ref{Table01},
we show values of
$|c(m)|^{2}$
and
$d(m)$
for
$m=0, 1, 2, 3, 4$.
Here, we look at the cases of $m=1$ and $m=3$.
When we let
$m=1$,
$|c(1)|^{2}$
is a small value and
$d(1)$
is nearly equal to unity.
Hence, if we put
$t=3\pi/(\sqrt{2}|g|)$,
we obtain the operation of the NS gate shown in Eq.~(\ref{definition-NS-gate-0})
with an upper bound for the error probability $0.138$.
When we let $m=3$,
$|c(3)|^{2}$ is nearly equal to zero and $d(3)$ is nearly equal to $-1$.
Thus, if we take $t=7\pi/(\sqrt{2}|g|)$,
we obtain
$\alpha|0\rangle_{\mbox{\scriptsize P}}
+
\beta|1\rangle_{\mbox{\scriptsize P}}
+
\gamma|2\rangle_{\mbox{\scriptsize P}}
\to
\alpha|0\rangle_{\mbox{\scriptsize P}}
-
\beta|1\rangle_{\mbox{\scriptsize P}}
-
\gamma|2\rangle_{\mbox{\scriptsize P}}$
with an upper bound for the error probability $0.0247$.
To turn over only the sign of the coefficient of $|1\rangle_{\mbox{\scriptsize P}}$,
we apply the phase shifter
$|n\rangle_{\mbox{\scriptsize P}}\to(-1)^{n}|n\rangle_{\mbox{\scriptsize P}}$
and
obtain an approximate NS gate.
From now on,
for the sake of simplicity,
we let $m=1$ through this paper.

\begin{figure}
\begin{center}
\includegraphics[scale=1.0]{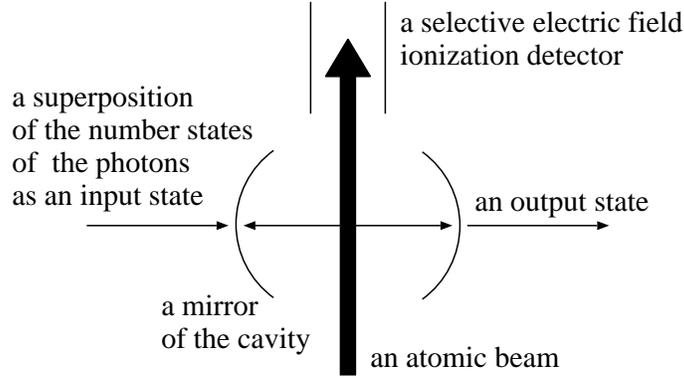}
\end{center}
\caption{An experimental setup for the NS gate realized by our scheme.
We provide a superposition of the number states of the photons as an input state
into the cavity from its left side.
It is reflected by mirrors of the cavity many times
and develops into a cavity field.
A slow beam of the two-level atom travels through the cavity and causes the Jaynes-Cummings interaction with the cavity field.
After the time evolution,
for the sake of simplicity,
we assume that the cavity field flies away from the cavity to its right side.
If we let the time of flight of the atom through the cavity be equal to
$T=3\pi/(\sqrt{2}|g|)$,
this implementation works as the NS gate approximately.
A selective electric field ionization detector makes a distinction between the atom's ground state $|g\rangle_{\mbox{\scriptsize A}}$
and excited state $|e\rangle_{\mbox{\scriptsize A}}$,
and we can examine whether the NS gate works or fails.}
\label{Figure01}
\end{figure}

In Fig.~\ref{Figure01},
we show an outline of an experimental setup for the above method that realizes the NS gate.
The important things are as follows.
We have to provide a superposition of the number states of the photons,
$|0\rangle_{\mbox{\scriptsize P}}$, $|1\rangle_{\mbox{\scriptsize P}}$, and $|2\rangle_{\mbox{\scriptsize P}}$,
inside the cavity.
Then,
we have to cause the Jaynes-Cummings interaction between the atom and the photons.

First, we provide a superposition of the number states of the photons into the cavity from its left side.
The photons are reflected by the mirrors of the cavity many times and they develop into the cavity mode.
We let the length between the mirrors
be equal to a half of the wavelength of the cavity mode.
Thus, the cavity mode forms a standing wave.

Second, we put the two-level atom at an anti-node of the standing wave of the cavity mode.
For example, we can capture an ionized atom in a certain region by the Paul trap (a quadruple ion trap)
\cite{Raizen1992,Roos1999}.
We can also put the atom in a certain area by injecting it as a slow beam.
Then, the cavity mode interacts with the atom as the JCM.
If we let the time of flight of the atom be equal to
$T=3\pi/(\sqrt{2}|g|)$, and if we observe $|g\rangle_{\mbox{\scriptsize A}}$ with the selective electric field detector in Fig.~\ref{Figure01},
we can realize an approximate NS gate.

If we can use a coherent state (a laser light) as the input state,
the experiment is achieved without problem.
In contrast, if we have to provide an arbitrary superposition of the number states of the photons as an input,
it is difficult to succeed in performing the experiment.
Details of experimental techniques are discussed in Ref.~\cite{Azuma2011}.

\section{\label{section-semiclassical-decoherence-model}
The semiclassical model causing the decoherence of the NS gate}
In this section,
we consider the semiclassical model that causes the decoherence for the NS gate.
In this model,
we introduce the interaction between the dipole moment of the two-level atom and the external time-varying electric field.
Because we let the external field vary stochastically,
the JCM suffers from the decoherence.
We examine the time evolution of the system with the Monte Carlo simulation.
This semiclassical treatment is inspired by Ref.~\cite{Palma1996}.
This approach prevents the system including many interesting quantum properties,
for example,
the qubit-environment entanglement.
However, we adopt this method to simplify calculations.

We let the dipole moment of the atom $e\hat{\mbox{\boldmath $r$}}$ interact with the external electric field $\mbox{\boldmath $E$}$.
The Hamiltonian of the system is given as follows \cite{Schleich2001}:
\begin{eqnarray}
H'
&=&
-e\hat{\mbox{\boldmath $r$}}\cdot\mbox{\boldmath $E$} \nonumber \\
&=&
(\hbar/2)|\kappa|\sigma_{y}E.
\label{Hamiltonian-electric-field-dipole-interaction}
\end{eqnarray}
We explain the derivation of the above Hamiltonian in Appendix~\ref{appendix-electric-field-dipole-interaction}.
In Eq.~(\ref{Hamiltonian-electric-field-dipole-interaction}),
$E$ denotes an amplitude of the electric field $\mbox{\boldmath $E$}$.
The definition of $\kappa$ is given in Appendix~\ref{appendix-electric-field-dipole-interaction} as well.
From now on,
for the sake of simplicity,
we describe $|\kappa|E$ as $E$.
Thus, the physical dimension of $E$ becomes the inverse of time.

Moreover,
we consider that $E(t)$ varies in time in a stochastic manner.
Thus,
we can write down the Hamiltonian as
\begin{equation}
H'(t)
=
(\hbar/2)
\sigma_{y}\tilde{E}(t).
\label{Hamiltonian-dash-definition-0}
\end{equation}
The tilde placed on top of $\tilde{E}(t)$ indicates
that it is a stochastic variable varying in time.

Here,
we assume that a unitary operator of the time evolution induced by the Hamiltonian $H'(t)$
for the finite time increment $\Delta t$ is given by
\begin{equation}
U'(\Delta t;t)
=
\exp[-i\Delta tH'(t)/\hbar].
\label{unitary-time-evolution-operator-0}
\end{equation}
Strictly speaking,
because $H'(t)$
depends on the time variable $t$,
the time evolution operator $U'(\Delta t;t)$ must not take the form of Eq.~(\ref{unitary-time-evolution-operator-0}).
To deal with this problem rigorously,
we have to solve the time-dependent Schr\"{o}dinger equation with the Hamiltonian that relies on the time variable.
However,
because $\tilde{E}(t)$ is the stochastic variable
and it is not an ordinary dynamical one,
we treat it in a semiclassical manner as shown in Eq.~(\ref{unitary-time-evolution-operator-0}).

Here, we define the stochastic variable $\tilde{E}(t)$ in concrete terms.
First,
we take a threshold of the probability $p$ between zero and $1/2$ as $0\leq p\leq 1/2$.
Second,
we select a random number $r$ uniformly distributed over the interval $[0,1]$
for each time step
and we obtain a sequence of random numbers as $\{r(\Delta t), r(2\Delta t), ..., r(n\Delta t)\}$,
where
we let $\Delta t$ be a very small discrete time step.
Third,
a sequence of the stochastic variable $\{\tilde{E}(0), \tilde{E}(\Delta t), \tilde{E}(2\Delta t), ..., \tilde{E}(n\Delta t)\}$
is given as follows:
\begin{equation}
\tilde{E}(0)=0,
\label{definition-stochastic-variable-0}
\end{equation}
\begin{equation}
\tilde{E}(t+\Delta t)
=
\left\{
\begin{array}{lll}
\tilde{E}(t)-\delta E & \quad & 0\leq r(t+\Delta t)<p, \\
\tilde{E}(t)          & \quad & p\leq r(t+\Delta t)< 1-p, \\
\tilde{E}(t)+\delta E & \quad & 1-p\leq r(t+\Delta t)\leq 1. \\
\end{array}
\right.
\label{definition-stochastic-variable-1}
\end{equation}
Thus,
to
generate the stochastic variable $\tilde{E}(t)$,
we need to prepare three constants,
$p$, $n(=t/\Delta t)$, and $\delta E$.
In Eq.~(\ref{definition-stochastic-variable-1}),
$\tilde{E}(t+\Delta t)$ depends on $\tilde{E}(t)$.
Thus, the classical stochastic field $\tilde{E}(t)$ is non-memoryless.
The time evolution of $\tilde{E}(t)$ depends on its past history.

Under the semiclassical treatment, approximate time evolution of the state at $t=n\Delta t$ is given by
\begin{eqnarray}
|\psi(n\Delta t)\rangle
&=&
[U_{\mbox{\scriptsize I}}(\Delta t)(U'(\Delta t;n\Delta t)\otimes\mbox{\boldmath $I$}_{\mbox{\scriptsize P}})]
[U_{\mbox{\scriptsize I}}(\Delta t)(U'(\Delta t;(n-1)\Delta t)\otimes\mbox{\boldmath $I$}_{\mbox{\scriptsize P}})]
... \nonumber \\
&&
\quad
\times
[U_{\mbox{\scriptsize I}}(\Delta t)(U'(\Delta t;2\Delta t)\otimes\mbox{\boldmath $I$}_{\mbox{\scriptsize P}})]
[U_{\mbox{\scriptsize I}}(\Delta t)(U'(\Delta t;\Delta t)\otimes\mbox{\boldmath $I$}_{\mbox{\scriptsize P}})]
|\psi(0)\rangle.
\label{time-evolution-state-vector-semiclassical-0}
\end{eqnarray}
In Eq.~(\ref{time-evolution-state-vector-semiclassical-0}),
$\mbox{\boldmath $I$}_{\mbox{\scriptsize P}}$ and $U_{\mbox{\scriptsize I}}(\Delta t)$ denote
the identity operator acting on the photons
and
the time evolution operator induced by $H_{\mbox{\scriptsize I}}$, respectively.
We pay attention to the fact that the operator $U'(\Delta t;t)$ acts only on the two-level atom.

The state vector $|\psi(n\Delta t)\rangle$ obtained in Eq.~(\ref{time-evolution-state-vector-semiclassical-0})
is just a single sample generated from the initial state
$|\psi(0)\rangle$
and a sequence of the stochastic variable
\\
$\{\tilde{E}(0), \tilde{E}(\Delta t), \tilde{E}(2\Delta t), ..., \tilde{E}(n\Delta t)\}$.
In other words,
we produce $|\psi(n\Delta t)\rangle$ from a single sequence of random numbers
$\{r(\Delta t), r(2\Delta t), ..., r(n\Delta t)\}$.
To obtain an expectation value of a physical quantity,
we have to generate many samples and perform the Monte Carlo simulation.
Thus,
we do not describe the system as the pure state $|\psi(n\Delta t)\rangle$
but as a mixed state,
\begin{equation}
\rho(n\Delta t)=(1/M)\sum_{m=1}^{M}|\psi(n\Delta t)\rangle_{m}{}_{m}\langle\psi(n\Delta t)|,
\label{stochastic-mixed-state-0}
\end{equation}
where $|\psi(n\Delta t)\rangle_{m}$ represents a wave function of the $m$th sample for $m=1, 2, ..., M$,
and $M$ represents the total number of samples.
From this ensemble averaging,
dephasing occurs in the NS gate.

In this semiclassical picture,
it is possible to consider a single member of the ensemble as a fully physical entity.
For instance,
it could describe a random interaction between the two-level atom and thermal radiation.
To obtain averages of physical quantities induced by random phenomena,
we have to collect many samples.

In this paper,
we focus on the fidelity and the Bloch vector as physical quantities that we obtain by numerical simulations.
We define the fidelity of the mixed state $\rho(n\Delta t)$
given by Eq.~(\ref{stochastic-mixed-state-0}) as follows \cite{Jozsa1994}.
We write the time-evolved state with $\tilde{E}(n\Delta t)=0$ $\forall n\geq 0$ as $|\psi(n\Delta t)\rangle_{0}$.
Thus,
$|\psi(n\Delta t)\rangle_{0}$ represents the time evolution of the wave function without decoherence.
In other words,
$|\psi(n\Delta t)\rangle_{0}$ develops only due to the Hamiltonian $H_{\mbox{\scriptsize I}}$
and $H'$ has no effect on it.
The Monte Carlo average of the fidelity is given by
\begin{eqnarray}
F(n\Delta t)
&=&
{}_{0}\langle\psi(n\Delta t)|\rho(n\Delta t)|\psi(n\Delta t)\rangle_{0} \nonumber \\
&=&
(1/M)\sum_{m=1}^{M}
|_{0}\langle\psi(n\Delta t)|\psi(n\Delta t)\rangle_{m}|^{2}.
\label{Monte-Carlo-fidelity-0}
\end{eqnarray}
The Monte Carlo average of the Bloch vector $\mbox{\boldmath $S$}(n\Delta t)$ is given by
\begin{eqnarray}
\rho_{\mbox{\scriptsize A}}(n\Delta t)
&=&
\mbox{Tr}_{\mbox{\scriptsize P}}[\rho(n\Delta t)] \nonumber \\
&=&
(1/2)[\mbox{\boldmath $I$}_{\mbox{\scriptsize A}}+\mbox{\boldmath $S$}(n\Delta t)\cdot\mbox{\boldmath $\sigma$}].
\label{Bloch-vector-Monte-Carlo-average-0}
\end{eqnarray}

The numerical simulation is carried out for $0\leq t\leq T$,
where $T=3\pi/(\sqrt{2}|g|)$ is the time  when the NS gate works.
We introduce the total number of time steps $N$, and we define $\Delta t=T/N$.
We study variance and a distribution of the stochastic variable $\tilde{E}(n\Delta t)$
in Appendix~\ref{appendix-stochastic-variable}.

\section{\label{section-numerical-simulation-semiclassical-model}
Numerical simulations of the semiclassical model}
In this section,
we investigate the decoherence of the semiclassical model introduced in Sect.~\ref{section-semiclassical-decoherence-model}
by numerical simulations.
Throughout this section,
we always use the following parameters for calculations unless we note otherwise.
First of all,
we define parameters of the Jaynes-Cummings interaction according to Ref.~\cite{Rempe1987}.
We consider $63p_{3/2}\leftrightarrow 61d_{5/2}$ transition of ${}^{85}\mbox{Rb}$.
The frequency and the wavelength of the atomic transition are given by
$f=21{\,}456.0\times 10^{6}$ Hz
and
$\lambda=1.397{\,}24\times 10^{-2}$ m, respectively.
The coupling constant is given by
$g=(1/70)\times 10^{6}$ $\mbox{s}^{-1}$.
We can estimate the time required for the operation of the NS gate at
$T=3\pi/(\sqrt{2}g)\simeq 4.67 \times 10^{-4}$ s.
We set the total number of time steps to $N=10^{5}$
and let the time increment for the temporal change of the stochastic variable $\tilde{E}(t)$
be equal to $\Delta t=T/N$.
We put $M=8\times 10^{5}$ for the total number of the Monte Carlo samples.
We use the Fortran 90 compiler with the double precision for carrying out numerical calculations.
We generate random numbers with the method of the Mersenne Twister using the free software MT19937.

Because $\Delta t=4.67\times 10^{-9}$ s,
we can estimate a frequency of $\tilde{E}(t)$ at around $2.14\times 10^{8}$ Hz.
This frequency is smaller than that of the atomic transition $f=2.15\times 10^{10}$ Hz.
Thus, putting $N=10^{5}$ is consistent with the rotating wave approximation that is applied to obtain the JCM.
Here, we make a remark about the Rabi oscillation induced by injecting a coherent light to the two-level atom.
We can estimate the frequency of the Rabi oscillation at around $g\sim 1.43\times 10^{4}$ Hz,
which is much smaller than that of $\tilde{E}(t)$.
However, this fact does not cause any problems to our model.
Because $g\propto\sqrt{\omega/V}$,
where $\omega$ and $V$ denote the angular frequency of the coherent light and the volume of the cavity respectively,
the coupling constant $g$ depends on the shape of the cavity.
We have to consider the stochastic noise $\tilde{E}(t)$
from the Rabi oscillation separately.
The inverse of the coupling constant $g$ characterizes the processing time of the NS gate as $T=3\pi/(\sqrt{2}|g|)$.
Thus, the processing time of the NS gate is comparable to a period of the Rabi oscillation.

In this section,
we let the initial states be given by
$(1/\sqrt{3})(|g,0\rangle+|g,1\rangle+|g,2\rangle)$,
$|0_{+}\rangle$,
and
$|g,1\rangle$,
and simulate their time evolution numerically.
To carry out numerical simulations actually,
we restrict the dimension of the Hilbert space to twelve and assume that its orthonormal basis is given by
\begin{equation}
\{|g,0\rangle, |g,1\rangle, ..., |g,5\rangle, |e,0\rangle, |e,1\rangle, ..., |e,5\rangle \}.
\label{twelve-basis-vectors-01}
\end{equation}

First,
we write down the time evolution of the twelve basis vectors caused by $U_{\mbox{\scriptsize I}}(\Delta t)$ as follows:
\begin{eqnarray}
U_{\mbox{\scriptsize I}}(\Delta t)|g,0\rangle
&=&
|g,0\rangle,
\label{time-evolution-g0} \\
U_{\mbox{\scriptsize I}}(\Delta t)|g,k\rangle
&=&
\cos(|g|\sqrt{k}\Delta t)|g,k\rangle
-
i(g/|g|)\sin(|g|\sqrt{k}\Delta t)|e,k-1\rangle \nonumber \\
&&\quad
\mbox{for $k=1, ..., 5$},
\label{time-evolution-gk} \\
U_{\mbox{\scriptsize I}}(\Delta t)|e,k\rangle
&=&
-i(g/|g|)\sin(|g|\sqrt{k+1}\Delta t)|g,k+1\rangle
+
\cos(|g|\sqrt{k+1}\Delta t)|e,k\rangle \nonumber \\
&&\quad
\mbox{for $k=0, ..., 4$},
\label{time-evolution-ek} \\
U_{\mbox{\scriptsize I}}(\Delta t)|e,5\rangle
&=&
\cos(|g|\sqrt{6}\Delta t)|e,5\rangle.
\label{time-evolution-e5}
\end{eqnarray}
Equation~(\ref{time-evolution-e5}) is an approximation of the following relation:
\begin{equation}
U_{\mbox{\scriptsize I}}(\Delta t)|e,5\rangle
=
-i(g/|g|)\sin(|g|\sqrt{6}\Delta t)|g,6\rangle
+
\cos(|g|\sqrt{6}\Delta t)|e,5\rangle.
\label{time-evolution-e5-org}
\end{equation}
Because we have to let the dimension of the Hilbert space be finite,
we adopt Eq.~(\ref{time-evolution-e5}) rather than Eq.~(\ref{time-evolution-e5-org}).
Second,
we describe matrix elements of
$U'(\Delta t;n\Delta t)\otimes\mbox{\boldmath $I$}_{\mbox{\scriptsize P}}$.
Because
\begin{equation}
U'(\Delta t;t)
=
\cos[\frac{\Delta t}{2}\tilde{E}(t)]\mbox{\boldmath $I$}_{\mbox{\scriptsize A}}
-i
\sin[\frac{\Delta t}{2}\tilde{E}(t)]\sigma_{y},
\end{equation}
we obtain
\begin{eqnarray}
\langle i|U'(\Delta t;n\Delta t)\otimes\mbox{\boldmath $I$}_{\mbox{\scriptsize P}}|i\rangle
&=&
\cos[\frac{\Delta t}{2}\tilde{E}(n\Delta t)]
\quad
\mbox{for $i\in\{|g,k\rangle,|e,k\rangle:k=0,1,...,5\}$}, \nonumber \\
\langle i|U'(\Delta t;n\Delta t)\otimes\mbox{\boldmath $I$}_{\mbox{\scriptsize P}}|j\rangle
&=&
-
\sin[\frac{\Delta t}{2}\tilde{E}(n\Delta t)] \nonumber \\
&&\quad
\mbox{for $i\in\{|e,k\rangle :k=0,1,...,5\}$,} \nonumber \\
&&\quad\quad\quad
\mbox{$j\in\{|g,l\rangle :l=0,1,...,5\}$}, \nonumber \\
\langle i|U'(\Delta t;n\Delta t)\otimes\mbox{\boldmath $I$}_{\mbox{\scriptsize P}}|j\rangle
&=&
\sin[\frac{\Delta t}{2}\tilde{E}(n\Delta t)] \nonumber \\
&&\quad
\mbox{for $i\in\{|g,k\rangle :k=0,1,...,5\}$,} \nonumber \\
&&\quad\quad\quad
\mbox{$j\in\{|e,l\rangle :l=0,1,...,5\}$}.
\label{matrix-element-Udash}
\end{eqnarray}

Here,
we define the initial state as
\begin{equation}
|\psi(0)\rangle =(1/\sqrt{3})(|g,0\rangle +|g,1\rangle +|g,2\rangle),
\label{initial-state-g0g1g2}
\end{equation}
and examine the time evolution of the Bloch vector of the two-level atom.
We calculate the Bloch vector as the ensemble average of the Monte Carlo simulation.
If we define the initial state as Eq.~(\ref{initial-state-g0g1g2}),
$S_{x}(t)=0$ $\forall t\geq 0$ holds for $\tilde{E}(t)=0$ $\forall t\geq 0$.
In other words,
if we think about the time evolution of Eq.~(\ref{initial-state-g0g1g2}) without decoherence,
$S_{x}(t)$ is always equal to zero.
Thus,
$S_{x}(t)$ is not a suitable physical quantity for examining the decoherence.
Hence,
from now on,
we compute the time variations of $S_{y}$, $S_{z}$, and $|\mbox{\boldmath $S$}|^{2}$,
and examine the decoherence of the system with them.

\begin{figure}
\begin{minipage}{0.48\hsize}
\begin{center}
\includegraphics[scale=0.72]{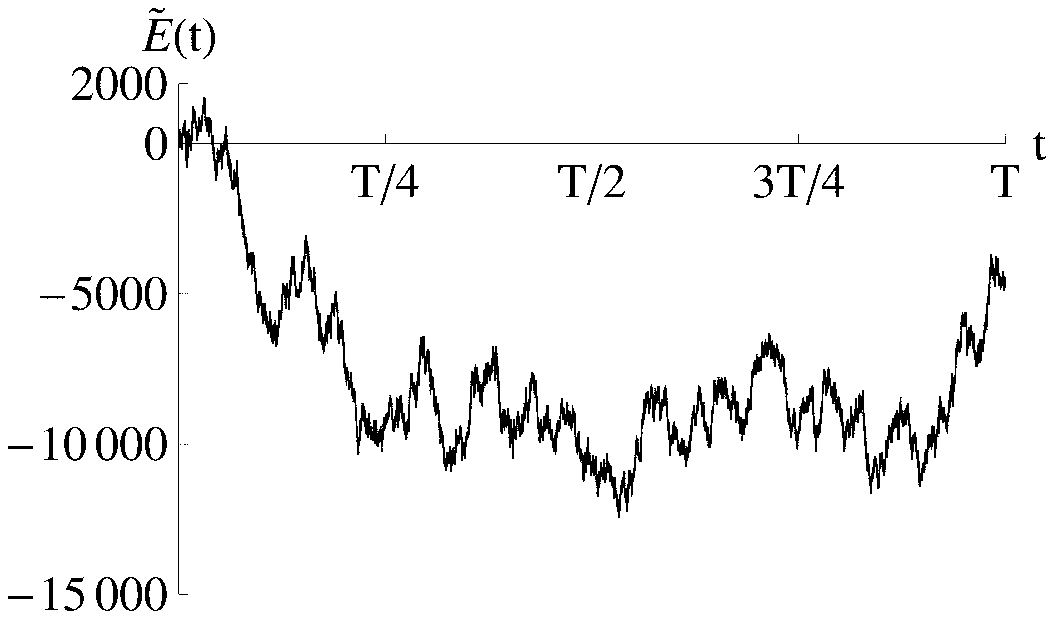}
\end{center}
\caption{A graph of the time variation for the external field of the stochastic variable $\tilde{E}(t)$
on a single sample with $\delta E=50.0$ and $p=0.2$.}
\label{Figure02}
\end{minipage}
\begin{minipage}{0.02\hsize}
\hspace{0cm}
\end{minipage}
\begin{minipage}{0.48\hsize}
\begin{center}
\includegraphics[scale=0.72]{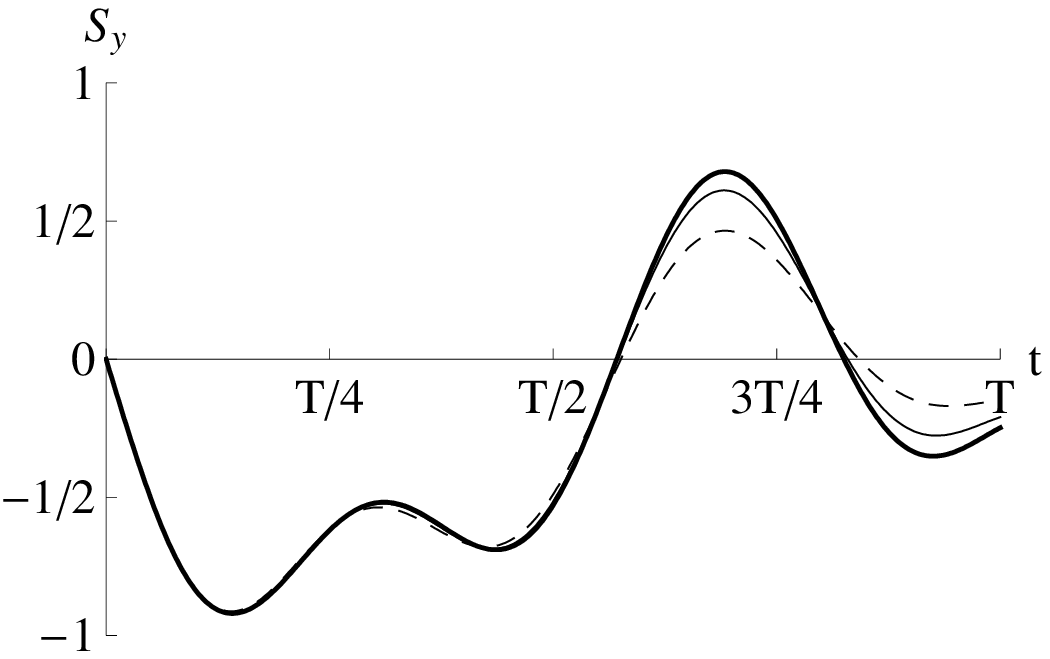}
\end{center}
\caption{Graphs of the time variations of $S_{y}$
with the initial state given by Eq.~(\ref{initial-state-g0g1g2}).
We put $p=0.2$.
A thick solid curve,
a thin solid curve,
and a thin dashed curve represent
$\delta E=0.0$ (no decoherence),
$50.0$,
and $100.0$, respectively.}
\label{Figure03}
\end{minipage}
\end{figure}

\begin{figure}
\begin{minipage}{0.48\hsize}
\begin{center}
\includegraphics[scale=0.72]{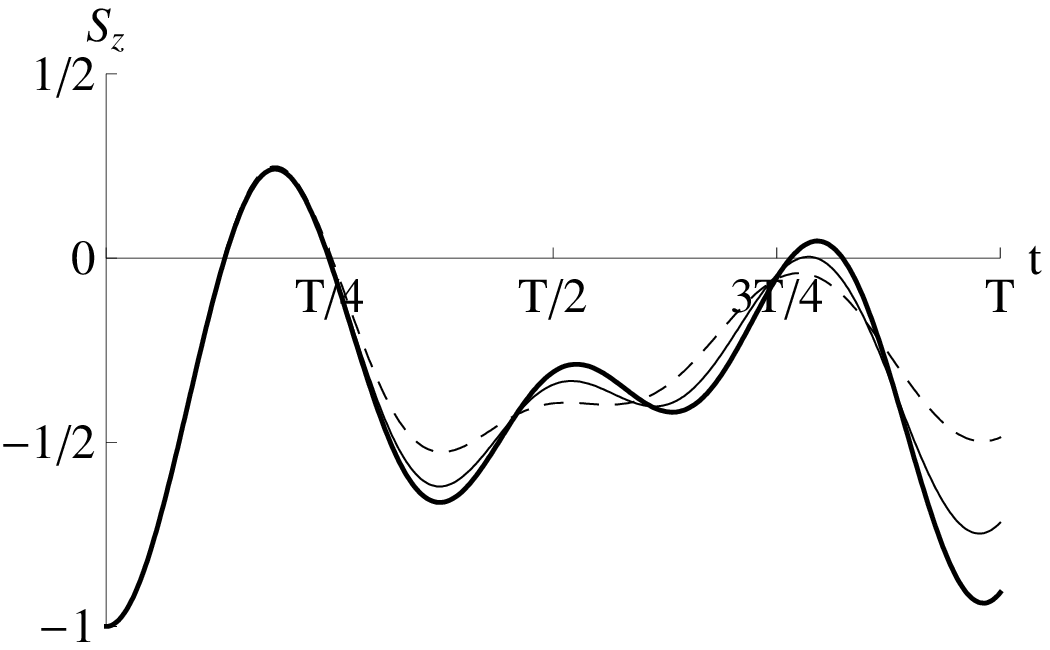}
\end{center}
\caption{Graphs of the time variations of $S_{z}$
with the initial state given by Eq.~(\ref{initial-state-g0g1g2}).
We put $p=0.2$.
A thick solid curve,
a thin solid curve,
and a thin dashed curve represent
$\delta E=0.0$ (no decoherence),
$50.0$,
and $100.0$, respectively.}
\label{Figure04}
\end{minipage}
\begin{minipage}{0.02\hsize}
\hspace{0cm}
\end{minipage}
\begin{minipage}{0.48\hsize}
\begin{center}
\includegraphics[scale=0.72]{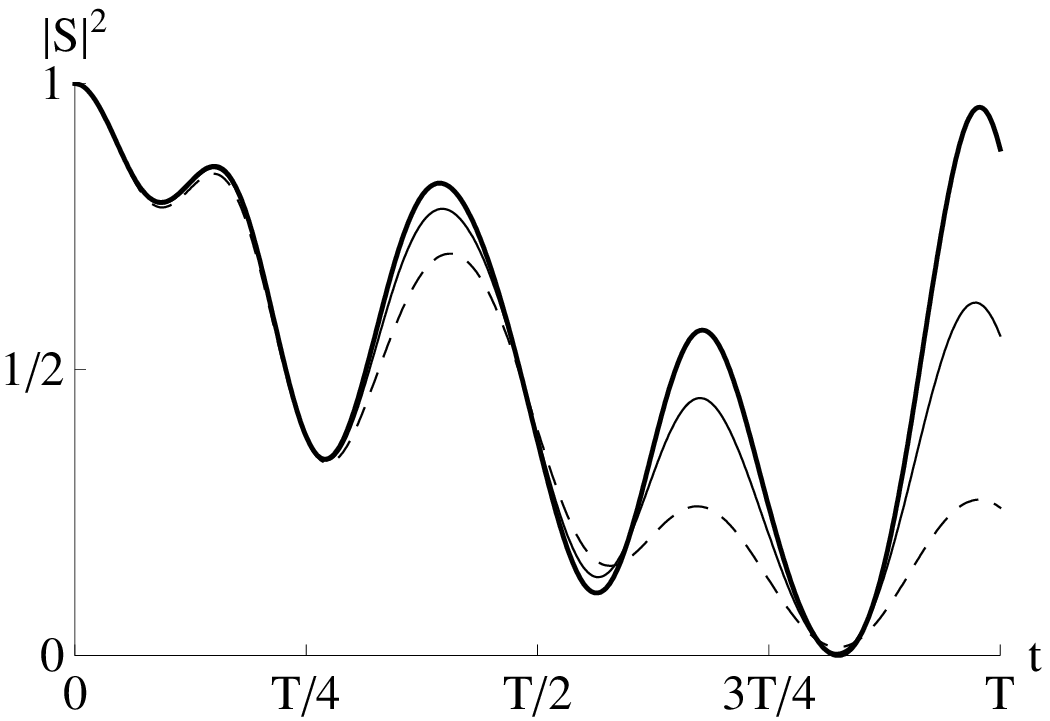}
\end{center}
\caption{Graphs of the time variations of $|\mbox{\boldmath $S$}|^{2}$
with the initial state given by Eq.~(\ref{initial-state-g0g1g2}).
We put $p=0.2$.
A thick solid curve,
a thin solid curve,
and a thin dashed curve represent
$\delta E=0.0$ (no decoherence),
$50.0$,
and $100.0$, respectively.}
\label{Figure05}
\end{minipage}
\end{figure}

Before we mention results of the Monte Carlo simulations,
we show a time variation of the stochastic field $\tilde{E}(t)$ for a single sample,
for example.
In Fig.~\ref{Figure02},
we plot it with $\delta E=50.0$ and $p=0.2$.

In Figs.~\ref{Figure03}, \ref{Figure04}, and \ref{Figure05},
we show time variations of $S_{y}$, $S_{z}$, and $|\mbox{\boldmath $S$}|^{2}$,
respectively.
We put $p=0.2$.
A thick solid curve,
a thin solid curve,
and a thin dashed curve represent
$\delta E=0.0$ (no decoherence),
$50.0$,
and $100.0$, respectively.

From Figs.~\ref{Figure03} and \ref{Figure04},
we can derive the following conclusion.
In general,
dephasing phenomena of the Bloch vector are classified into two types.
The first one is called the $\mbox{T}_{1}$ decay,
which is the relaxation of the $z$-component $S_{z}(t)$.
The second one is called the $\mbox{T}_{2}$ decay,
which is the relaxation of the $x$- and $y$-components $S_{x}(t)\mbox{\boldmath $e$}_{x}+S_{y}(t)\mbox{\boldmath $e$}_{y}$.
In Figs.~\ref{Figure03} and \ref{Figure04},
we can observe both the $\mbox{T}_{1}$ and $\mbox{T}_{2}$ decays.
(In the semiclassical model of Ref.~\cite{Palma1996},
only the $T_{2}$ decay occurs.)

\begin{table}
\caption{A table of $S_{z}$ at $t=T[=3\pi/(\sqrt{2}g)]$ for various numbers of the Monte Carlo samples $M$.
We put $\delta E=100.0$ and $p=0.2$.
The number of samples varies as $M=4\times 10^{5}$, $8\times 10^{5}$, and $1.2\times 10^{6}$.
From this table, we cam conclude that the physical quantities have three significant figures in the simulation.}
\label{Table02}
\begin{center}
\begin{tabular}{|c|c|} \hline
$M$ & $S_{z}(t=T)$ \\
\hline
$4\times 10^{5}$   & $-0.487{\,}217{\,}6$ \\
$8\times 10^{5}$   & $-0.486{\,}847{\,}7$ \\
$1.2\times 10^{6}$ & $-0.487{\,}201{\,}9$ \\
\hline
\end{tabular}
\end{center}
\end{table}

In Table~\ref{Table02},
we examine the number of significant figures for the physical quantities obtained by the simulations.
We give the averages of $S_{z}(t=T)$ for various numbers of the Monte Carlo samples
as $M=4\times 10^{5}$, $8\times 10^{5}$, and $1.2\times 10^{6}$ in Table~\ref{Table02}.
From these results,
we understand that the number of significant figures for physical quantities is equal to three.

Because we let the dimension of the Hilbert space be finite as shown in Eq.~(\ref{twelve-basis-vectors-01})
and adopt Eq.~(\ref{time-evolution-e5}) rather than Eq.~(\ref{time-evolution-e5-org}),
the norm of the state vector is not conserved.
If we put $\delta E=100.0$, $p=0.2$,
the total number of time steps $N=10^{5}$,
and the total number of the Monte Carlo samples $M=8\times 10^{5}$,
we obtain
\begin{equation}
1-\parallel |\psi(T)\rangle\parallel^{2}\simeq 4.66\times 10^{-6}.
\end{equation}
Thus, we can consider that the conservation of the norm of $|\psi(T)\rangle$ holds well approximately.

\begin{figure}
\begin{minipage}{0.48\hsize}
\begin{center}
\includegraphics[scale=0.72]{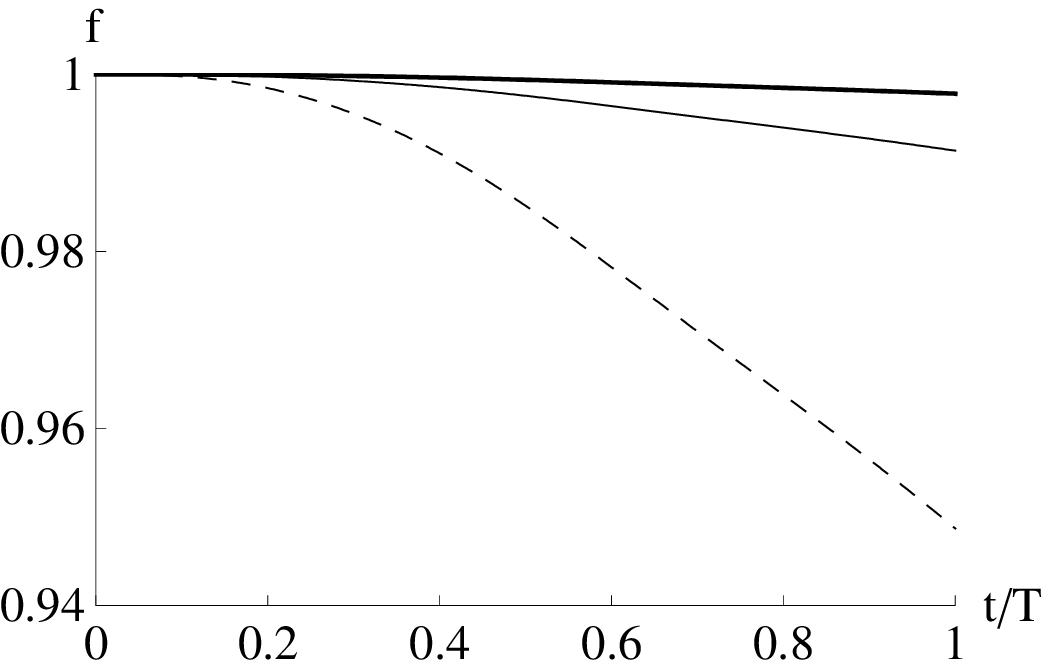}
\end{center}
\caption{Graphs of time variations of the fidelity with the initial state $|\psi(0)\rangle=|0_{+}\rangle$.
The horizontal and vertical axes represent $t/T$ and $F$, respectively.
We put $p=0.1$.
A thick solid curve, a thin solid curve, and a thin dashed curve
represent
$\delta E=5.0$,
$10.0$, and $25.0$, respectively.}
\label{Figure06}
\end{minipage}
\begin{minipage}{0.02\hsize}
\hspace{0cm}
\end{minipage}
\begin{minipage}{0.48\hsize}
\begin{center}
\includegraphics[scale=0.72]{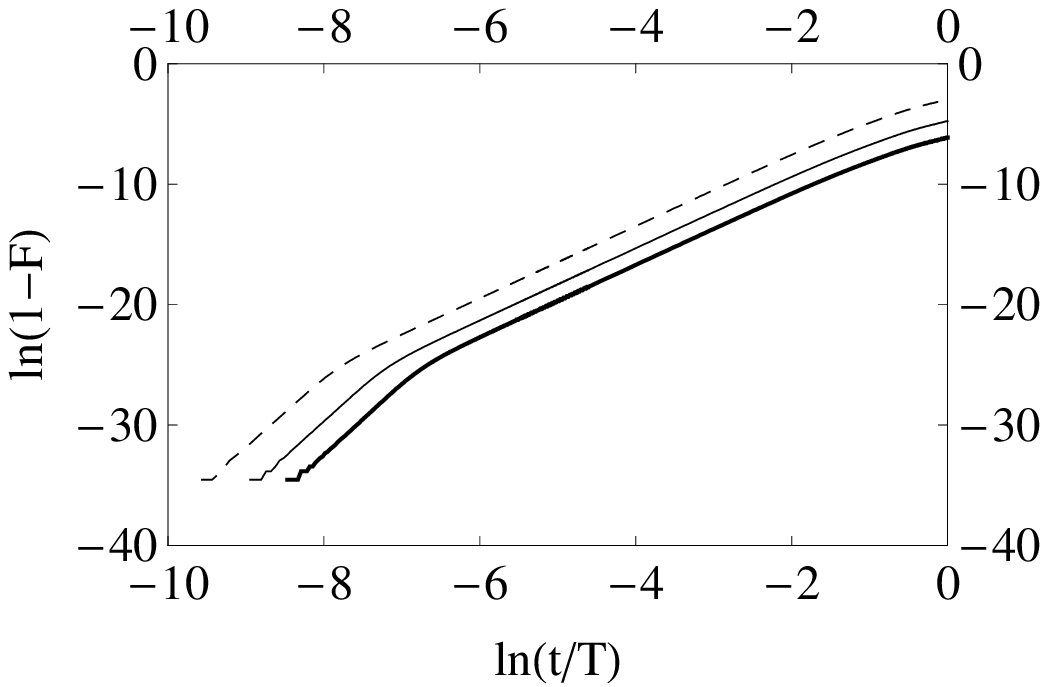}
\end{center}
\caption{Graphs of time variations of the fidelity with the initial state $|\psi(0)\rangle=|0_{+}\rangle$.
The horizontal and vertical axes represent $\ln(t/T)$ and $\ln(1-F)$, respectively.
We put $p=0.1$.
A thick solid curve, a thin solid curve, and a thin dashed curve
represent
$\delta E=5.0$,
$10.0$, and $25.0$, respectively.}
\label{Figure07}
\end{minipage}
\end{figure}

Next,
we
investigate the time variation of the fidelity.
In Fig.~\ref{Figure06},
we show time variations of the fidelity $F$ with preparing the initial state as $|\psi(0)\rangle=|0_{+}\rangle$.
We express the time variable as $t/T$,
and it becomes dimensionless.
In Fig.~\ref{Figure06},
we put $p=0.1$.
A thick solid curve,
a thin solid curve,
and a thin dashed curve represent
$\delta E=5.0$,
$10.0$,
and $25.0$, respectively.

\begin{table}
\caption{Results obtained by fitting the curves of Fig.~\ref{Figure07}
with linear functions given by Eq.~(\ref{fidelity-linear-fitting-0}) in the range of
$0.002\leq t/T\leq 0.05$,
namely
$-6.22\leq\ln(t/T)\leq -3.00$.}
\label{Table03}
\begin{center}
\begin{tabular}{|c|ccc|} \hline
$i$ & $\delta E$ & $a_{i}$ & $b_{i}$ \\
\hline
$0$ & $5.0$  & $-4.73$ & $2.99$ \\
$1$ & $20.0$ & $-3.33$ & $3.00$ \\
$2$ & $25.0$ & $-1.50$ & $3.00$ \\
\hline
\end{tabular}
\end{center}
\end{table}

We redraw the graphs of Fig.~\ref{Figure06} in Fig.~\ref{Figure07}
with letting the horizontal and vertical axes represent $\ln(t/T)$ and $\ln(1-F)$, respectively.
Looking at Fig.~\ref{Figure07},
we notice that we can fit curves with linear functions in the range of $0.002\leq t/T\leq 0.05$,
that is to say $-6.22\leq\ln(t/T)\leq -3.00$.
The linear functions obtained from fitting are given by
\begin{equation}
\ln(1-F)\simeq a_{i}+b_{i}\ln(t/T),
\label{fidelity-linear-fitting-0}
\end{equation}
where constants $\{a_{i}\}$ and $\{b_{i}\}$ are shown in Table~\ref{Table03}.
The indices $i=0$, $1$, and $2$ denote $\delta E=5.0$, $10.0$, and $25.0$, respectively.

\begin{figure}
\begin{center}
\includegraphics[scale=0.72]{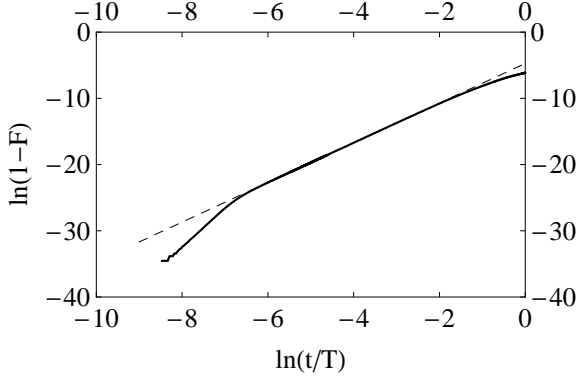}
\end{center}
\caption{Graphs of the time variation of the fidelity and its fitted line.
The horizontal and vertical axes represent $\ln(t/T)$ and $\ln(1-F)$, respectively.
A thick solid curve represents
the time variation of the fidelity with $|\psi(0)\rangle=|0_{+}\rangle$,
$\delta E=5.0$, and $p=0.1$.
A thin dashed line represents the fitted line that we obtain from the curve of $\ln(1-F)$
within the range of $-6.22\leq\ln(t/T)\leq -3.00$.}
\label{Figure08}
\end{figure}

In Fig.~\ref{Figure08},
we redraw the time variation of the fidelity in Fig.~\ref{Figure07} with $\delta E=5.0$ and $p=0.1$.
We also plot the linear function,
with which we fit the curve of $\ln(1-F)$ within the range of $-6.22\leq\ln(t/T)\leq -3.00$.
In Fig.~\ref{Figure08},
a thick solid curve represents $\ln(1-F)$ and a thin dashed line represents the linear function obtained from fitting.
From discussion in Sect.~\ref{section-perturbation-theory-fidelity},
we conclude that we can fit a graph of $\ln(1-F)$ plotted against $\ln(t/T)$
with a linear function in the range of $\Delta t/(2p)\ll t\ll 1/g$.
Figure~\ref{Figure08} certifies this result.
[We can rewrite the condition
$\Delta t/(2p)\ll t\ll 1/g$
as
$5.0\times10^{-5}\ll t/T \ll 0.150$
or
$-9.90\ll \ln(t/T)\ll -1.90$ explicitly.]

Looking at Table~\ref{Table03},
we notice that $\{b_{i}\}$ do not change depending on $\delta E$.
By contrast, because
\begin{eqnarray}
a_{2}-a_{0}
&=&
3.23,
\quad\quad
2(\ln 25.0 - \ln 5.0)
\simeq
3.22, \nonumber \\
a_{1}-a_{0}
&=&
1.39,
\quad\quad
2(\ln 10.0 - \ln 5.0)
\simeq
1.39,
\end{eqnarray}
we can suppose
\begin{equation}
a_{i}=\mbox{Const.}+2\ln(\delta E).
\label{coefficient-approximation-a}
\end{equation}
Examining how $a$ and $b$ depend on $p$, $N$, and $g$ by numerical simulations in the same manner shown above,
we can suppose the following relations:
\begin{eqnarray}
\ln(1-F)&\simeq&a+b\ln(t/T), \nonumber \\
a&=&1.98+2\ln(\delta E)+\ln p+\ln N -2\ln g, \nonumber \\
b&=&2.99.
\label{fidelity-linear-fitting-0+}
\end{eqnarray}

\begin{figure}
\begin{minipage}{0.48\hsize}
\begin{center}
\includegraphics[scale=0.72]{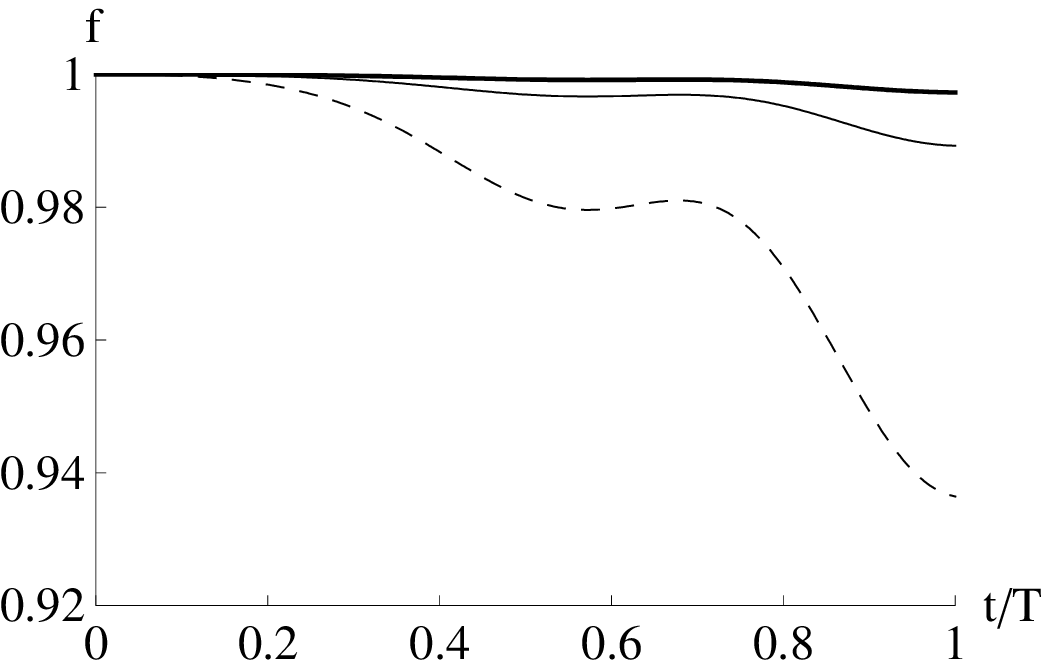}
\end{center}
\caption{
Graphs of time variations of the fidelity with the initial state $|\psi(0)\rangle=|g,1\rangle$.
The horizontal and vertical axes represent $t/T$ and $F$, respectively.
We put $p=0.1$.
A thick solid curve, a thin solid curve, and a thin dashed curve
represent
$\delta E=5.0$,
$10.0$, and $25.0$, respectively.}
\label{Figure09}
\end{minipage}
\begin{minipage}{0.02\hsize}
\hspace{0cm}
\end{minipage}
\begin{minipage}{0.48\hsize}
\begin{center}
\includegraphics[scale=0.72]{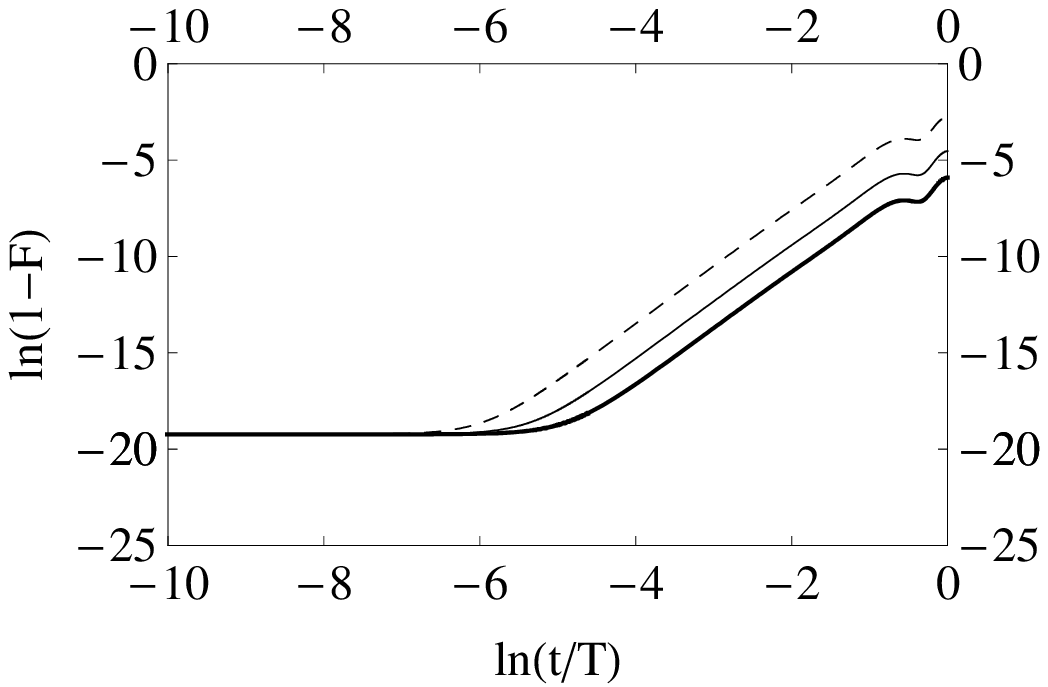}
\end{center}
\caption{Graphs of time variations of the fidelity with the initial state $|\psi(0)\rangle=|g,1\rangle$.
The horizontal and vertical axes represent $\ln(t/T)$ and $\ln(1-F)$, respectively.
We put $p=0.1$.
A thick solid curve, a thin solid curve, and a thin dashed curve
represent
$\delta E=5.0$,
$10.0$, and $25.0$, respectively.}
\label{Figure10}
\end{minipage}
\end{figure}

\begin{table}
\caption{Results obtained by fitting the curves of Fig.~\ref{Figure10}
with linear functions given by Eq.~(\ref{fidelity-linear-fitting-0}) in the range of
$0.02\leq t/T\leq 0.1$,
namely $-3.91\leq\ln(t/T)\leq -2.30$.}
\label{Table04}
\begin{center}
\begin{tabular}{|c|ccc|} \hline
$i$ & $\delta E$ & $a_{i}$ & $b_{i}$ \\
\hline
$0$ & $5.0$  & $-4.89$ & $2.94$ \\
$1$ & $10.0$ & $-3.47$ & $2.96$ \\
$2$ & $25.0$ & $-1.62$ & $2.96$ \\
\hline
\end{tabular}
\end{center}
\end{table}

In Fig.~\ref{Figure09},
we plot time variations of the fidelity with preparing the initial state as
$|\psi(0)\rangle
=
|g,1\rangle$.
We put $p=0.1$.
A thick solid curve, a thin solid curve, and a thin dashed curve represent
$\delta E=5.0$, $10.0$, and $25.0$, respectively.
In Fig.~\ref{Figure10},
we redraw the graphs of Fig.~\ref{Figure09} with setting the horizontal and vertical axes
to $\ln(t/T)$ and $\ln(1-F)$, respectively.
Looking at Fig.~\ref{Figure10},
we notice that linear functions can approximate to the graphs within the range of $0.02\leq t/T\leq 0.1$,
that is to say $-3.91\leq\ln(t/T)\leq -2.30$.
We can write the approximate linear functions in the form of Eq.~(\ref{fidelity-linear-fitting-0}).
The constants $\{a_{i}\}$ and $\{b_{i}\}$ are given in Table~\ref{Table04}.
The indices $i=0$, $1$, and $2$ denote $\delta E=5.0$, $10.0$, and $25.0$, respectively.

Looking at Table~\ref{Table04},
we notice that $\{b_{i}\}$ do not change depending on $\delta E$.
In contrast, because
\begin{eqnarray}
a_{2}-a_{0}
&=&
3.11,
\quad\quad
2(\ln 25.0 - \ln 5.0)
\simeq
3.22, \nonumber \\
a_{1}-a_{0}
&=&
1.26,
\quad\quad
2(\ln 10.0 - \ln 5.0)
\simeq
1.39,
\label{a0-a1-a2-relations-0}
\end{eqnarray}
we can suppose that Eq.~(\ref{coefficient-approximation-a}) holds.
Examining how $a$ and $b$ depend on $p$, $N$, and $g$ by numerical simulations,
we can suppose the following relations:
\begin{eqnarray}
\ln(1-F)&\simeq&a+b\ln(t/T), \nonumber \\
a&=&1.98+2\ln(\delta E)+\ln p+\ln N -2\ln g, \nonumber \\
b&=&2.94.
\label{linear-fitting-g1}
\end{eqnarray}

\begin{figure}
\begin{center}
\includegraphics[scale=0.72]{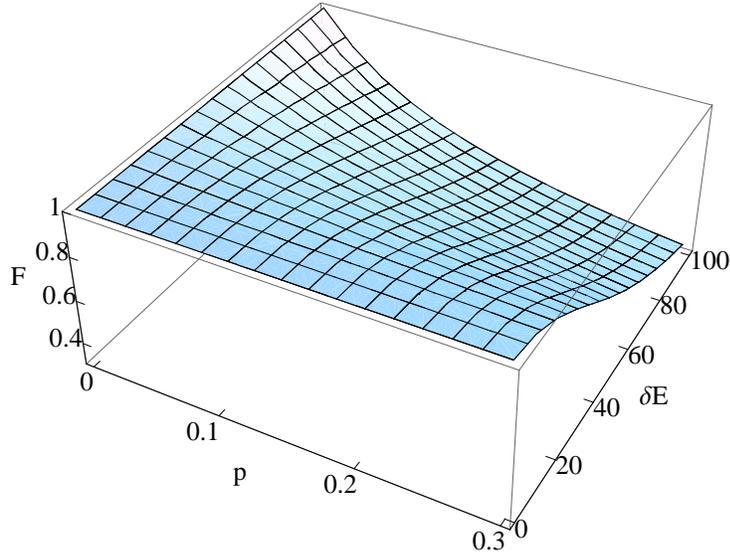}
\end{center}
\caption{A three-dimensional surface plot of $F(T)$
on the $p$-$\delta E$ plane with the initial state given by Eq.~(\ref{initial-state-g0g1g2}).
The surface is plotted within the range of $0.0\leq p\leq 0.3$ and $0.0\leq\delta E\leq 100.0$.
To generate a mesh on the $p$-$\delta E$ plane,
we divide the intervals of $0.0\leq p\leq 0.3$ and $0.0\leq\delta E\leq 100.0$
into twelve and ten equal segments, respectively.
Thus,
the lengths of the line segments of $p$ and $\delta E$ are equal to $0.025$ and $10.0$,
respectively.
To create the three-dimensional surface,
the Monte Carlo simulations are carried out on $120$ vertices on the mesh in total.}
\label{Figure11}
\end{figure}

In Fig.~\ref{Figure11},
we show a three-dimensional surface plot of $F(T)$,
the fidelity at time $t=T$,
on the $p$-$\delta E$ plane with the initial state given by Eq.~(\ref{initial-state-g0g1g2}).
We examine how $F(T)$ depends on $p$ and $\delta E$
within the range of $0.0\leq p\leq 0.3$ and $0.0\leq\delta E\leq 100.0$.
Looking at Fig.~\ref{Figure11},
we notice that $F(T)$ is convex downward for $p$ around $p=0.0$.
For example,
we can observe $\partial^{2}F/\partial p^{2}>0$ for $p\simeq 0.0$ and $\delta E=100.0$.
Contrastingly,
$F(T)$ is convex upward for $\delta E$ around $\delta E=0.0$.
For example,
we can observe $\partial^{2}F/(\partial \delta E)^{2}<0$ for $\delta E\simeq 0.0$ and $p=0.3$.

\section{\label{section-perturbation-theory-fidelity}
Evaluation of the fidelity with the time-dependent perturbation theory
for the stochastic process}
In this section,
we evaluate the fidelity with the time-dependent perturbation theory
for the stochastic process.
We write the initial state
and the state at time $T=N\Delta t$ as $|\psi(0)\rangle$ and $|\psi(T)\rangle$,
respectively.
According to the semiclassical model introduced in Sect.~\ref{section-numerical-simulation-semiclassical-model},
we can write $|\psi(T)\rangle$ in the form
\begin{equation}
|\psi(T)\rangle
=
[\exp(-i\frac{\Delta t}{\hbar}H_{\mbox{\scriptsize I}})
\exp(-i\frac{\Delta t}{\hbar}H')]^{N}|\psi(0)\rangle,
\label{state-vector-semiclassical-model-0}
\end{equation}
where $H_{\mbox{\scriptsize I}}$ and $H'$ are given by Eqs.~(\ref{Hamiltonian-definition-0}) and (\ref{Hamiltonian-dash-definition-0}),
respectively.
Here,
we neglect time dependence of $H'$ in Eq.~(\ref{state-vector-semiclassical-model-0}).
Using the Lie-Trotter product formula \cite{Trotter1959,Kato1978,Reed1980,Wiebe2010},
\begin{eqnarray}
\lim_{N\to\infty}
(e^{itA/N}e^{itB/N})^{N}
&=&
\lim_{N\to\infty}[e^{it(A+B)}+O(\frac{t^{2}}{N})] \nonumber \\
&=&
e^{it(A+B)}
\quad\quad
\forall t>0,
\end{eqnarray}
we can approximate $|\psi(T)\rangle=|\psi(N\Delta t)\rangle$ in the large $N$ limit to
\begin{equation}
|\psi(T)\rangle
\simeq
\exp[-i\frac{T}{\hbar}(H_{\mbox{\scriptsize I}}+H')]|\psi(0)\rangle.
\end{equation}
Thus,
regarding $H_{\mbox{\scriptsize I}}$ and $H'$ as the unperturbed and perturbing Hamiltonians respectively,
and letting $H=H_{\mbox{\scriptsize I}}+H'$ be the total Hamiltonian,
we evaluate $|\psi(t)\rangle$ with the time-dependent perturbation theory.

We give the eigenstates and eigenvalues of $H_{\mbox{\scriptsize I}}$
in Eqs.~(\ref{eigen-system-2-0}) and (\ref{eigen-system-2-1}).
We set the initial state to $|\psi(0)\rangle=|0_{+}\rangle$.
According to the time-dependent perturbation theory up to the second order,
$|\psi(t)\rangle$ is given as follows \cite{Schiff1968}:
\begin{eqnarray}
|\psi(t)\rangle
&=&
\sum_{i\in\{(g,0), 0_{\pm}, 1_{\pm}, ...\}}c_{i}(t)\exp(-iE_{i}t/\hbar)|i\rangle, \nonumber \\
c_{i}(t)
&=&
c_{i,0_{+}}^{(0)}(t)+c_{i,0_{+}}^{(1)}(t)+c_{i,0_{+}}^{(2)}(t), \nonumber \\
c_{i,j}^{(0)}(t)
&=&
\delta_{ij}, \nonumber \\
c_{i,j}^{(1)}(t)
&=&
-\langle i|H'|j\rangle
\frac{\exp[i(E_{i}-E_{j})t/\hbar]-1}{E_{i}-E_{j}}, \nonumber \\
c_{i,j}^{(2)}(t)
&=&
-\frac{1}{i\hbar}\sum_{k\in\{(g,0), 0_{\pm}, 1_{\pm}, ...\}}
\langle i|H'|k\rangle\langle k|H'|j\rangle
\int_{0}^{t}\frac{e^{i(E_{i}-E_{j})t'/\hbar}-e^{i(E_{i}-E_{k})t'/\hbar}}{E_{k}-E_{j}}dt'.
\label{perturbation-theory-up-to-second-order-0}
\end{eqnarray}
In the derivation of Eq.~(\ref{perturbation-theory-up-to-second-order-0}),
we assume that $H'$ does not rely on the time variable $t$.
The index $j$ of $c_{i,j}^{(n)}(t)$ appearing in Eq.~(\ref{perturbation-theory-up-to-second-order-0})
implies that the initial state is given by $|\psi(0)\rangle=|j\rangle$.

Strictly speaking,
because $H'$ includes the stochastic variable $\tilde{E}(t)$,
it depends on the time variable $t$.
However,
because $\tilde{E}(t)$ is not an ordinary dynamical variable,
we interpret it as a variable that is independent of time $t$.
At the last stage of computing the fidelity,
we take an average of the stochastic variable $\tilde{E}(t)$.

We write the evolved state with $\tilde{E}(t)=0$ $\forall t\geq 0$,
that is to say the evolved state without decoherence as
\begin{equation}
|\psi(t)\rangle_{0}
=
\exp(-igt)|0_{+}\rangle.
\end{equation}
Then, up to the second order perturbation,
the fidelity is given as follows:
\begin{eqnarray}
F(t)
&=&
|{}_{0}\langle\psi(t)|\psi(t)\rangle|^{2} \nonumber \\
&\simeq&
|c^{(0)}_{0_{+},0_{+}}(t)+c^{(1)}_{0_{+},0_{+}}(t)+c^{(2)}_{0_{+},0_{+}}(t)|^{2}.
\end{eqnarray}
Clearly, we obtain $c^{(0)}_{0_{+},0_{+}}(t)=1$.
Because we can derive
$\langle 0_{+}|H'|0_{+}\rangle=0$ from Eqs.~(\ref{eigen-system-2-0}) and (\ref{Hamiltonian-dash-definition-0}),
we obtain $c^{(1)}_{0_{+},0_{+}}(t)=0$.
Next, we compute products of the matrix elements $\langle 0_{+}|H'|i\rangle\langle i|H'|0_{+}\rangle$
as follows:
\begin{eqnarray}
\langle 0_{+}|H'|i\rangle\langle i|H'|0_{+}\rangle
&=&
0
\quad\quad
\mbox{for $j\neq (g,0), 1_{\pm}$}, \nonumber \\
\langle 0_{+}|H'|g,0\rangle\langle g,0|H'|0_{+}\rangle
&=&
(1/8)\hbar^{2}\tilde{E}^{2}(t), \nonumber \\
\langle 0_{+}|H'|1_{\pm}\rangle\langle 1_{\pm}|H'|0_{+}\rangle
&=&
(1/16)\hbar^{2}\tilde{E}^{2}(t).
\end{eqnarray}
Thus,
we obtain
\begin{eqnarray}
c^{(2)}_{0_{+},0_{+}}(t)
&=&
-i
\frac{\tilde{E}^{2}(t)}{g}
\Biggl[
\frac{1}{8}
\Biggl(
t-\frac{1}{ig}(e^{igt}-1)
\Biggr) \nonumber \\
&&\quad
+\frac{1}{16(1-\sqrt{2})}
\Biggl(
t-\frac{1}{ig(1-\sqrt{2})}
(e^{i(1-\sqrt{2})gt}-1)
\Biggr) \nonumber \\
&&\quad
+\frac{1}{16(\sqrt{2}+1)}
\Biggl(
t-\frac{1}{ig(1+\sqrt{2})}
(e^{i(1+\sqrt{2})gt}-1)
\Biggr)
\Biggr].
\end{eqnarray}
Considering the limit $gt\ll 1$,
we obtain
\begin{equation}
c^{(2)}_{0_{+},0_{+}}(t)
\simeq
-\frac{1}{8}\tilde{E}^{2}(t)t^{2}.
\label{c-2-0p-0p}
\end{equation}

Putting the above results together,
we obtain the following conclusion.
If we put $|\psi(0)\rangle=|0_{+}\rangle$ on condition that $t\ll 1/g$,
we obtain
\begin{equation}
F(t)
\simeq
1-\frac{1}{4}\tilde{E}^{2}(t)t^{2},
\label{Fidelity-evaluation-0}
\end{equation}
where we assume $0\leq\tilde{E}^{2}(t)t^{2}\ll 1$.
Here,
we take an average of $\tilde{E}^{2}(t)$.
From Eqs.~(\ref{variance-original}) and (\ref{variance-0}),
assuming $\Delta t/(2p)\ll t$,
we obtain
\begin{equation}
\langle\tilde{E}^{2}(t)\rangle
=
2(\delta E)^{2}p\frac{t}{\Delta t},
\end{equation}
so that we achieve
\begin{equation}
F(t)
=
1-\frac{1}{2}(\delta E)^{2}p\frac{t^{3}}{\Delta t}.
\end{equation}

If we put $p=0.1$, $\delta E=5.0$, and $t=0.05T$,
we obtain
\begin{equation}
\langle\tilde{E}^{2}(t)\rangle t^{2}
\simeq
1.36\times 10^{-5}
\ll
1.
\end{equation}
At the same time, as mentioned in Sect.~\ref{section-numerical-simulation-semiclassical-model},
because of $\Delta t/(2pT)=5.0\times 10^{-5}$
and
$1/(gT)\simeq0.150$,
$\Delta t/(2p)\ll t\ll 1/g$ holds for $t=0.05T$.
Thus, our approximate calculation in Eq.~(\ref{Fidelity-evaluation-0}) is valid

Finally,
to let the time variable be dimensionless,
we rewrite $F(t)$ as $F(t/T)$ as follows:
\begin{equation}
F(t/T)
=
1-\frac{1}{2}(\delta E)^{2}p\frac{T^{3}}{\Delta t}\Biggl(\frac{t}{T}\Biggr)^{3}.
\label{Fidelity-evaluation-1}
\end{equation}
Because
\begin{equation}
T/\Delta t=N,
\quad\quad
T=\frac{3\pi}{\sqrt{2}g},
\end{equation}
we arrive at
\begin{equation}
F(t/T)
=
1-\frac{9\pi^{2}}{4}(\delta E)^{2}pN\frac{1}{g^{2}}\Biggl(\frac{t}{T}\Biggr)^{3},
\label{fidelity-form-1-1}
\end{equation}
\begin{equation}
9\pi^{2}/4\simeq 22.2,
\quad\quad
\ln(22.2)\simeq 3.10.
\label{fidelity-form-1-2}
\end{equation}
Equations (\ref{fidelity-form-1-1}) and (\ref{fidelity-form-1-2}) imply
\begin{equation}
\ln(1-F)
\simeq
3.10+2\ln(\delta E)+\ln p+\ln N-2\ln g+3\ln(t/T),
\label{fidelity-form-1-3}
\end{equation}
and Eq.~(\ref{fidelity-form-1-3}) is similar to Eq.~(\ref{fidelity-linear-fitting-0+}).
Specially,
Eq.~(\ref{fidelity-form-1-3}) explains the reason why the constant $b$ is nearly equal to three in Eq.~(\ref{fidelity-linear-fitting-0+}).

So far,
we have discussed the perturbation theory for the stochastic process with the initial state
$|\psi(0)\rangle
=
|0_{+}\rangle$
on condition that $\Delta t/(2p)\ll t\ll 1/g$.
Next,
we investigate the time evolution of $|\psi(t)\rangle$
with the initial state
$|\psi(0)\rangle=|g,1\rangle=(1/\sqrt{2})(|0_{+}\rangle-|0_{-}\rangle)$.

Because the evolved state without decoherence is given by
\begin{equation}
|\psi(t)\rangle_{0}
=
(1/\sqrt{2})
(
e^{-igt}|0_{+}\rangle
-
e^{igt}|0_{-}\rangle
),
\end{equation}
we can write down the fidelity as
\begin{equation}
F(t)
=
(1/2)
\left|
[
e^{igt}\langle 0_{+}|\psi(t)\rangle
-
e^{-igt}\langle 0_{-}|\psi(t)\rangle
]
\right|^{2}.
\end{equation}
Now,
we introduce the following notation.
We describe the time evolution of
$|\psi(0)\rangle=|0_{+}\rangle$
as
$|\psi(t);0_{+}\rangle$.
In a similar way,
we describe the time evolution of
$|\psi(0)\rangle=|0_{-}\rangle$
as
$|\psi(t);0_{-}\rangle$.
Then, we obtain
\begin{equation}
|\psi(t)\rangle
=
(1/\sqrt{2})
(
|\psi(t);0_{+}\rangle
-
|\psi(t);0_{-}\rangle
).
\end{equation}
Up to the second order perturbation,
because of the previous results,
we obtain the following relation with ease on condition that $gt\ll 1$:
\begin{eqnarray}
\langle 0_{+}|e^{igt}|\psi(t);0_{+}\rangle
&\simeq&
c^{(0)}_{0_{+},0_{+}}(t)
+
c^{(1)}_{0_{+},0_{+}}(t)
+
c^{(2)}_{0_{+},0_{+}}(t) \nonumber \\
&\simeq&
1-\frac{1}{8}\tilde{E}^{2}(t)t^{2}.
\end{eqnarray}
Carrying out similar calculations up to the second order perturbation for $gt\ll 1$,
we obtain
\begin{eqnarray}
\langle 0_{+}|e^{igt}|\psi(t);0_{-}\rangle
&\simeq&
0, \nonumber \\
\langle 0_{-}|e^{-igt}|\psi(t);0_{+}\rangle
&\simeq&
0, \nonumber \\
\langle 0_{-}|e^{-igt}|\psi(t);0_{-}\rangle
&\simeq&
1-\frac{1}{8}\tilde{E}^{2}(t)t^{2}.
\end{eqnarray}
Thus, we obtain the fidelity as Eqs.~(\ref{Fidelity-evaluation-0}), (\ref{Fidelity-evaluation-1}), and (\ref{fidelity-form-1-3})
for $\Delta t/(2p)\ll t\ll 1/g$.
Equation~(\ref{fidelity-form-1-3}) is similar to Eq.~(\ref{linear-fitting-g1}).
Specially, Eq.~(\ref{fidelity-form-1-3}) explains the reason why the constant $b$ is nearly equal to three in Eq.~(\ref{linear-fitting-g1}).

\section{\label{section-discussion}Discussion}
In this paper,
we investigate the decoherence of KLM's NS gate implemented with the JCM
in a semiclassical manner by introducing
the stochastic variable as the external electric field.
In this model,
we observe both the $T_{1}$ and $T_{2}$ decays.

In the semiclassical model and the quantum mechanical perturbative analysis for the stochastic process,
as results of Eqs.~(\ref{fidelity-linear-fitting-0+}),
(\ref{linear-fitting-g1}),
and (\ref{fidelity-form-1-3}),
we obtain the fidelity in the form
\begin{equation}
F(t)
=
1-\mbox{Const.}(\delta E)^{2}pN\frac{1}{g^{2}}
\Biggl(
\frac{t}{T}
\Biggr)^{3},
\end{equation}
for $T/(2pN)\ll t\ll 1/g$,
where $|\psi(0)\rangle=|0_{+}\rangle$ and $|g,1\rangle$.
We can expect this relation to be a useful formula for interpreting experimental data.

We can write down the Schr\"{o}dinger equation with the stochastic external field $\tilde{E}(t)$ as
\begin{equation}
i\hbar
\frac{\partial}{\partial t}
|\psi(t)\rangle
=
(H_{\mbox{\scriptsize I}}+
\frac{\hbar}{2}
\sigma_{y}
\tilde{E}(t))
|\psi(t)\rangle.
\label{Schrodinger-equation-stochastic}
\end{equation}
We can regard this equation as a stochastic differential equation.
Moreover,
Eq.~(\ref{Schrodinger-equation-stochastic}) is similar to the Langevin equation.
Thus,
we can expect to obtain a new example of the fluctuation-dispersion theorem.
This problem remains to be solved in future.

\appendix
\section{\label{appendix-electric-field-dipole-interaction}
The electric field-dipole interaction}
In this section, we formulate the electric field-dipole interaction \cite{Schleich2001}.
First of all,
we consider a hydrogen atom with a proton of mass $m_{\mbox{\scriptsize p}}$ at position $\mbox{\boldmath $r$}_{\mbox{\scriptsize p}}$
and an electron of mass $m_{\mbox{\scriptsize e}}$ at position $\mbox{\boldmath $r$}_{\mbox{\scriptsize e}}$.
Then, we define a dipole moment of the atom as follows:
\begin{eqnarray}
\hat{\mbox{\boldmath $P$}}
&\equiv&
e\hat{\mbox{\boldmath $r$}} \nonumber \\
&=&
e(\mbox{\boldmath $r$}_{\mbox{\scriptsize e}}-\mbox{\boldmath $r$}_{\mbox{\scriptsize p}}).
\end{eqnarray}
We assume that the atom is put in the electric field.
Moreover,
we assume that the electric field
does not change considerably over the size of the atom.
Thus, we obtain
\begin{equation}
\mbox{\boldmath $E$}(\mbox{\boldmath $r$}_{\mbox{\scriptsize e}},t)
\simeq
\mbox{\boldmath $E$}(\mbox{\boldmath $R$},t),
\end{equation}
where $\mbox{\boldmath $R$}$ is the centre-of-mass,
that is to say
\begin{equation}
\mbox{\boldmath $R$}
\equiv
(
m_{\mbox{\scriptsize e}}\mbox{\boldmath $r$}_{\mbox{\scriptsize e}}
+
m_{\mbox{\scriptsize p}}\mbox{\boldmath $r$}_{\mbox{\scriptsize p}}
)
/
(m_{\mbox{\scriptsize e}}+m_{\mbox{\scriptsize p}}).
\end{equation}
We can write down a potential energy of the dipole moment in the electric field as
\begin{eqnarray}
H'
&=&
-\hat{\mbox{\boldmath $P$}}\cdot\mbox{\boldmath $E$}(\mbox{\boldmath $R$},t) \nonumber \\
&=&
-e\hat{\mbox{\boldmath $r$}}\cdot\mbox{\boldmath $E$}(\mbox{\boldmath $R$},t).
\end{eqnarray}
Now, we treat the dipole moment $\hat{\mbox{\boldmath $P$}}$ in a quantum mechanical manner and regard
$\mbox{\boldmath $E$}(\mbox{\boldmath $R$},t)$ as the classical electric field.
Hence, we solve a problem concerning the Hamiltonian $H'$ semiclassically.

Here, we examine how to express the position operator $\hat{\mbox{\boldmath $r$}}$
with the eigenstates of the two-level atom
$\{|g\rangle_{\mbox{\scriptsize A}},|e\rangle_{\mbox{\scriptsize A}}\}$.
Because the eigenstates of wave functions
$\{\psi_{\mbox{\scriptsize A},g}(\mbox{\boldmath $r$}),
\psi_{\mbox{\scriptsize A},e}(\mbox{\boldmath $r$})\}$
have a well-defined parity,
that is to say both
$|\psi_{\mbox{\scriptsize A},g}(\mbox{\boldmath $r$})|^{2}$ and $|\psi_{\mbox{\scriptsize A},e}(\mbox{\boldmath $r$})|^{2}$
are symmetric functions for $\mbox{\boldmath $r$}$,
diagonal elements vanish as
\begin{eqnarray}
{}_{\mbox{\scriptsize A}}\langle j|\hat{\mbox{\boldmath $r$}}|j\rangle_{\mbox{\scriptsize A}}
&=&
\int d^{3}r
\;
|\psi_{\mbox{\scriptsize A},j}(\mbox{\boldmath $r$})|^{2}\mbox{\boldmath $r$} \nonumber \\
&=&
0
\quad\quad
\mbox{for $j\in\{g,e\}$}.
\end{eqnarray}
We can describe off-diagonal elements as follows:
\begin{eqnarray}
e{}_{\mbox{\scriptsize A}}\langle e|\hat{\mbox{\boldmath $r$}}|g\rangle_{\mbox{\scriptsize A}}
&=&
e
\int d^{3}r
\;
\psi_{\mbox{\scriptsize A},e}^{*}(\mbox{\boldmath $r$})
\mbox{\boldmath $r$}
\psi_{\mbox{\scriptsize A},g}(\mbox{\boldmath $r$}) \nonumber \\
&\equiv&
\mbox{\boldmath $P$}, \nonumber \\
e{}_{\mbox{\scriptsize A}}\langle g|\hat{\mbox{\boldmath $r$}}|e\rangle_{\mbox{\scriptsize A}}
&\equiv&
\mbox{\boldmath $P$}^{*}.
\end{eqnarray}
Thus, we can write down the dipole operator as
\begin{eqnarray}
e
\hat{\mbox{\boldmath $r$}}
&=&
e
(|g\rangle_{\mbox{\scriptsize A}}{}_{\mbox{\scriptsize A}}\langle g|
+
|e\rangle_{\mbox{\scriptsize A}}{}_{\mbox{\scriptsize A}}\langle e|)
\hat{\mbox{\boldmath $r$}}
(|g\rangle_{\mbox{\scriptsize A}}{}_{\mbox{\scriptsize A}}\langle g|
+
|e\rangle_{\mbox{\scriptsize A}}{}_{\mbox{\scriptsize A}}\langle e|) \nonumber \\
&=&
\mbox{\boldmath $P$}|e\rangle_{\mbox{\scriptsize A}}{}_{\mbox{\scriptsize A}}\langle g|
+
\mbox{\boldmath $P$}^{*}|g\rangle_{\mbox{\scriptsize A}}{}_{\mbox{\scriptsize A}}\langle e| \nonumber \\
&=&
\mbox{\boldmath $P$}\sigma_{+}+\mbox{\boldmath $P$}^{*}\sigma_{-}.
\end{eqnarray}

Hence, the Hamiltonian of the potential energy caused by the electric field-dipole interaction is given by
\begin{equation}
H'
=
-(\mbox{\boldmath $P$}\sigma_{+}+\mbox{\boldmath $P$}^{*}\sigma_{-})\cdot\mbox{\boldmath $E$}.
\end{equation}
Here, we introduce the following notation:
\begin{equation}
\mbox{\boldmath $E$}
=
E
\mbox{\boldmath $u$},
\end{equation}
where $E$ denotes an amplitude of the electric field and $\mbox{\boldmath $u$}$ represents a real unit vector.
Then, we can rewrite the Hamiltonian as
\begin{equation}
H'=-(\hbar/2)
(\kappa\sigma_{+}+\kappa^{*}\sigma_{-})E,
\end{equation}
where
\begin{equation}
\kappa=(2/\hbar)\mbox{\boldmath $P$}\cdot\mbox{\boldmath $u$}.
\end{equation}
Now, we introduce a phase as
\begin{equation}
\kappa=|\kappa|e^{i\varphi},
\end{equation}
and we rewrite the Hamiltonian as
\begin{equation}
H'=-(\hbar/2)
(e^{i\varphi}\sigma_{+}+e^{-i\varphi}\sigma_{-})|\kappa|E.
\end{equation}
Then, choosing the phase $\varphi=\pi/2$,
we arrive at
\begin{equation}
H'=(\hbar/2)
|\kappa|\sigma_{y}E.
\end{equation}

\section{\label{appendix-stochastic-variable}
Variance and a probability distribution of the stochastic variable $\tilde{E}(t)$}
In this section,
we study a time variation of the stochastic variable $\tilde{E}(t)$,
that is to say $\{\tilde{E}(0), \tilde{E}(\Delta t), ..., \tilde{E}(n\Delta t)\}$.
The definition of the stochastic variable $\tilde{E}(t)$ is given by Eqs.~(\ref{definition-stochastic-variable-0})
and
(\ref{definition-stochastic-variable-1}).
To compute $\tilde{E}(t)$,
we need three parameters, $p$, $n(=t/\Delta t)$, and $\delta E$.
We pay attention to the fact that $\tilde{E}(t)$ does not depend on $\Delta t$.
Here,
we investigate variance and a probability distribution of $\tilde{E}(t)$
in the Monte Carlo simulation.

Taking $M$ samples in total,
we obtain $M$ sequences of $(n+1)$ numbers,
\begin{equation}
\{\tilde{E}^{(m)}(0), \tilde{E}^{(m)}(\Delta t), ..., \tilde{E}^{(m)}(n\Delta t)\}
\quad\quad
\mbox{for $m=1, 2, ..., M$}.
\end{equation}
Clearly,
the following relation holds:
\begin{equation}
\langle\tilde{E}(n\Delta t)\rangle
=
\lim_{M\to\infty}
(1/M)\sum_{m=1}^{M}\tilde{E}^{(m)}(n\Delta t)=0
\quad\quad
\mbox{for $\forall n$}.
\end{equation}

\begin{figure}
\begin{minipage}{0.48\hsize}
\begin{center}
\includegraphics[scale=0.72]{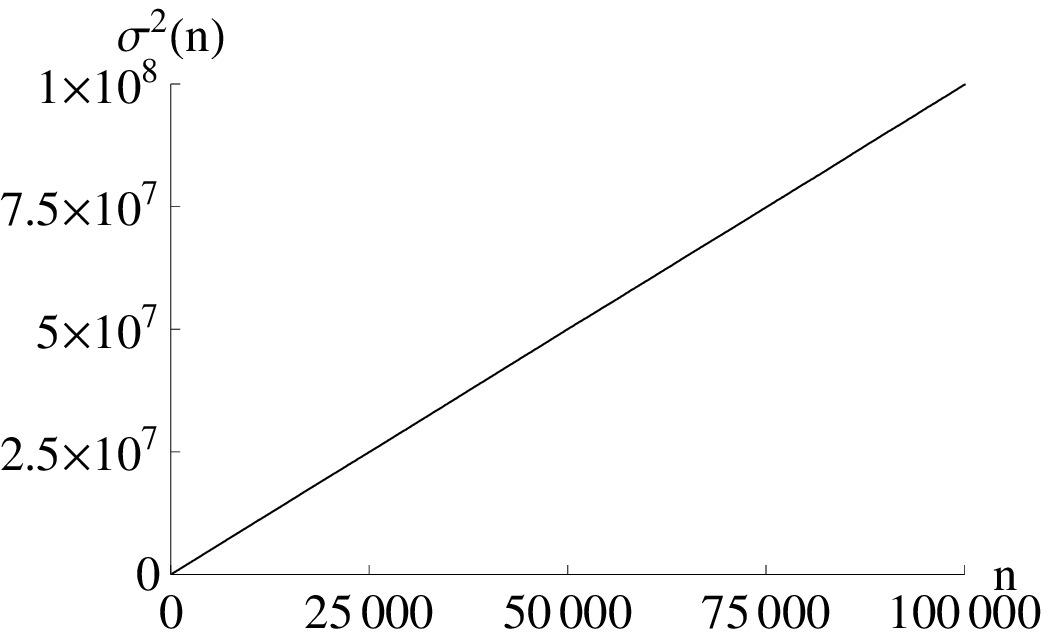}
\end{center}
\caption{A graph of $\sigma^{2}(n)$ for $1\leq n\leq 10^{5}$
with $p=0.2$ and $\delta E=50.0$.
We set the number of samples for the Monte Carlo simulation to $M=8\times 10^{5}$.
The graph approximates to a linear function that passes through the origin.}
\label{Figure12}
\end{minipage}
\begin{minipage}{0.02\hsize}
\hspace{0cm}
\end{minipage}
\begin{minipage}{0.48\hsize}
\begin{center}
\includegraphics[scale=0.72]{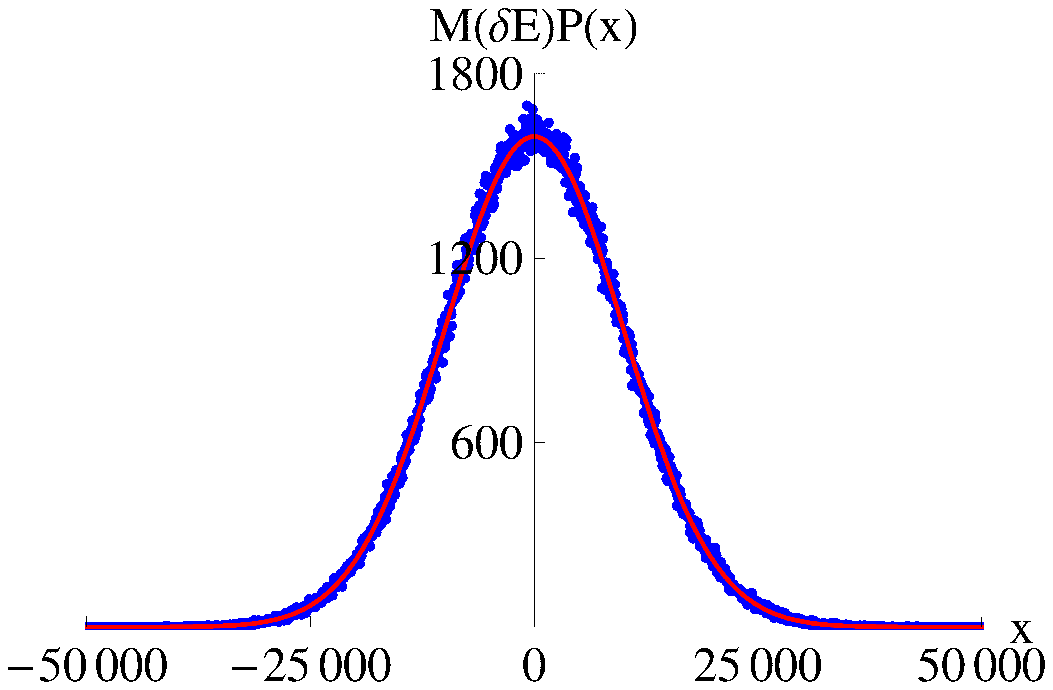}
\end{center}
\caption{Blue dots represent a distribution of $\{\tilde{E}^{(m)}(n\Delta t):m=1, 2, ..., M\}$
with $p=0.2$, $n=10^{5}$, and $\delta E=50.0$.
We put $M=8\times 10^{5}$.
The horizontal axis $x$ represents the value of $\tilde{E}^{(m)}(n\Delta t)$.
The vertical axis represents the number of $\{\tilde{E}^{(m)}(n\Delta t)\}$,
each of which is equal to $x$.
A red curve represents a normal distribution of $M(\delta E)P(x)$, where $P(x)$ is
given by Eq.~(\ref{normal-distribution}).}
\label{Figure13}
\end{minipage}
\end{figure}

Next,
we define the variance of $\tilde{E}(t)$
for the Monte Carlo simulation as follows:
\begin{eqnarray}
\sigma^{2}(n)
&=&
\langle\tilde{E}^{2}(n\Delta t)\rangle \nonumber \\
&=&
\lim_{M\to\infty}
(1/M)
\sum_{m=1}^{M}
\tilde{E}^{(m)}(n\Delta t)^{2}.
\label{variance-original}
\end{eqnarray}
In Fig.~\ref{Figure12},
we plot $\sigma^{2}(n)$ for $1\leq n\leq 10^{5}$
with $p=0.2$ and $\delta E=50.0$.
We set the total number of samples to $M=8\times 10^{5}$.
A graph of Fig.~\ref{Figure12} approximates to a linear function that passes through the origin.
From this numerical result,
we can suppose the following relation:
\begin{equation}
\sigma^{2}(n)=2(\delta E)^{2}pn.
\label{variance-0}
\end{equation}
From similar numerical calculations,
we can suppose an explicit form of $\langle\tilde{E}^{4}(n\Delta t)\rangle$ as
\begin{equation}
\langle\tilde{E}^{4}(n\Delta t)\rangle=12(\delta E)^{4}p^{2}n^{2}.
\end{equation}
Putting these considerations together,
we can expect the following relations:
\begin{equation}
\left\{
\begin{array}{lll}
\langle\tilde{E}^{l}(n\Delta t)\rangle=0 & \quad & \mbox{for $l=1, 3, ...$ ($l$: odd)}, \\
\langle\tilde{E}^{l}(n\Delta t)\rangle\propto n^{l/2} & \quad & \mbox{for $l=2, 4, ...$ ($l$: even)}. \\
\end{array}
\right.
\label{average-E-l-0}
\end{equation}

Next,
for a certain fixed $n$,
we
plot a distribution of $\{\tilde{E}^{(m)}(n\Delta t): m=1, 2, ..., M\}$ with blue dots in Fig.~\ref{Figure13}.
Looking at Fig.~\ref{Figure13},
we notice that a graph of blue dots approximates to a normal distribution drawn as a red curve.
The probability $P(x)$ that $\tilde{E}(n\Delta t)$ is equal to $x$ at the $n$th step is given by
\begin{equation}
P(x)
=
\frac{1}{\sqrt{2\pi}\sigma(n)}
\exp
[-\frac{x^{2}}{2\sigma^{2}(n)}],
\label{normal-distribution}
\end{equation}
where $\sigma^{2}(n)$ represents the variance given by Eq.~(\ref{variance-0}).
The number of $\{\tilde{E}^{(m)}(n\Delta t)\}$, each of which is equal to $x$,
is given by $M(\delta E)P(x)$.

Here,
we pay attention to the following facts.
The distribution of $\{\tilde{E}^{(m)}(n\Delta t):m=1, 2, ..., M\}$
for a certain fixed $n$ approximates to the normal distribution well on condition that $n$ becomes large enough.
Here,
we estimate $n$ which lets the distribution of $\{\tilde{E}^{(m)}(n\Delta t):m=1, 2, ..., M\}$ be close to the normal distribution well.
We have derived $\sigma^{2}(n)=2(\delta E)^{2}pn$ already in Eq.(\ref{variance-0}).
However,
$\tilde{E}^{(m)}(n\Delta t)$ varies in a discretized manner with a discrete unit $\delta E$.
Thus,
if $\sigma^{2}(n)$ is larger than $(\delta E)^{2}$ enough,
that is to say in case of $2pn\gg 1$,
the distribution of $\tilde{E}(n\Delta t)$ approximates the normal distribution well.
Thus,
on condition that $n\gg 1/(2p)$,
we can approximate $\tilde{E}(n\Delta t)$ to the normal distribution.

\end{document}